\documentclass[a4paper,11pt]{article}

\usepackage[a4paper,
  left=2.5cm, right=2.5cm,
  top= 3cm, bottom=4cm]{geometry}
\usepackage{amsmath}
\usepackage{oldgerm}
\usepackage{amssymb}
\usepackage{bbm}
\usepackage[dvips]{graphicx}
\usepackage{epsfig}
\usepackage{color}
\usepackage{cite}
\usepackage{epic}
\usepackage{hyperref}

\usepackage{empheq}

\numberwithin{equation}{section}

\usepackage[utf8]{inputenc}

\usepackage{graphicx}
\usepackage{mathrsfs}
\usepackage{amsmath}
\usepackage{amssymb}
\usepackage{braket} 
\usepackage{hyperref}
\usepackage{frontespizio}
\usepackage{comment}
\usepackage{epsfig}
\usepackage{float}
\usepackage{color}
\usepackage[title]{appendix}
\usepackage{multirow}
\usepackage{tikz}
\usetikzlibrary{decorations.pathmorphing}
\usepackage{varwidth}
\usepackage{tikz}
\usetikzlibrary{backgrounds,patterns,calc}
\usepackage{xparse}
\usepackage{caption}
\usepackage{enumitem}
\setlist[itemize]{leftmargin=*}

\usetikzlibrary{decorations.pathmorphing}

\tikzset{zigzag/.style={decorate, decoration=zigzag}}

\newcommand{\citep}{\cite}

\newcommand{\bea}{\begin{eqnarray}}
\newcommand{\eea}{\end{eqnarray}}
\newcommand{\be}{\begin{equation}}
\newcommand{\ee}{\end{equation}}
\newcommand{\ba}{\begin{align}}
\newcommand{\ea}{\end{align}}

\usepackage{subfigure}

\DeclareMathOperator{\sech}{sech}

\newcommand{\mpl}{{M_{\mathrm{Pl}}}}

\def\0{{\boldsymbol 0}}

\def\k{{\vec{k}}}

\def\x{{\vec{x}}}

\title{}
\author{}

\numberwithin{equation}{section}

\begin{document}
%
%
\begin{titlepage}

\vspace{-0.5cm} {\flushright {\small{IFT-UAM/CSIC-21-157}}} \\
\vspace{1cm}
\begin{center}
{\huge\bf{
Density Perturbations and Primordial Non-Gaussianities in a Closed Universe\\[0.3cm]}}

\vspace{1.2truecm}

{\fontsize{10.5}{18}\selectfont
{\bf Sebasti\'an C\'espedes,${}^{\rm a}$ Senarath de Alwis,${}^{\rm b}$ Francesco Muia ${}^{\rm c}$ and Fernando Quevedo${}^{\rm c}$
}}
\vspace{.5truecm}

{\small{\it $^{a}$ Instituto de F\'{i}sica Te\'orica UAM-CSIC
C/ Nicol\'as Cabrera 13-15, \\ Campus de Cantoblanco, 28049 Madrid, Spain}}  \\ 
  {\small {\it $^{b}$ Physics Department, University of Colorado, Boulder, CO 80309 USA}}\\
{\small {\it $^{c}$ DAMTP,  Centre for Mathematical Sciences,  University of Cambridge,\\ Wilberforce Road,  Cambridge, CB3 0WA, UK}}

  \vskip 2.2cm

 \begin{abstract}
{\small{The spatial curvature of the universe is not  yet known. Even though at present the Universe is very close to being essentially flat and most signatures of  curvature appear to  have been diluted by inflation, if the number of e-foldings during inflation is close to the minimum necessary to explain the horizon problem, the curvature of the universe may have left imprints in the cosmic microwave background (CMB) that may be observable, especially at large angles. Motivated by general results on quantum cosmology and  using  effective field theory techniques, we develop a general approach for analytically computing the power spectrum of density perturbations for a closed universe. Following a Hamiltonian formalism we determine the corresponding Bunch-Davis vacuum, find analytic expressions for two-point functions and higher correlators, expanding in terms of $S^3$ harmonics. In particular we concentrate on potential implications for observable  non-Gaussianities. We consider cubic  interactions as well as higher derivative ones to explore the consequence of a speed of sound $c_s\neq 1$. For large multipoles curvature effects are negligible and reproduce the known results for the flat case. However, they depart from the flat space result for relatively small multipoles. In this limit non-Gaussianities may lead to potentially observable values of $f_{\rm NL}$. In particular we find terms in $f_{\rm NL}$ that are absent in the flat space case, which are important at large scales and may be observable even if a long period of inflation dilutes the curvature. We compare our results with previous discussions in the literature.}}
 
 \end{abstract}
\end{center}

\end{titlepage}






\newpage

\tableofcontents

\renewcommand*{\thefootnote}{\arabic{footnote}}
\setcounter{footnote}{0}

\newpage

\section{Introduction}
One of the most important parameters of the homogeneous Friedman-Lemaitre-Robinson-Walker (FLRW) cosmological scenario is the spatial curvature of the universe $k=0,\pm 1$, which provides a model independent contribution to the energy density $\Omega_k=-k/(aH)^2$ for  a scale factor $a$ and Hubble parameter $H$. Current observations favour an almost flat universe with the Planck data 2018~\cite{Planck:2018jri} reporting $\Omega_k=-0.0007\pm 0.0019$. Combined with BAO observations this lead to a universe consistent with exactly flat but with a margin of error slightly preferring a closed universe. The statistical significance and error bar of these results is a question of current debate
 (see for instance \cite{Handley:2019tkm,DiValentino:2020hov} and  \cite{Efstathiou:2020wem} for  opposing views). As further observations improve with time, the parameter range for $\Omega_k$ could be determined with higher precision. 

Determining the value of $k$ is of crucial importance to understand the early and late stages of the universe. For the early stages, simple solvable models of quantum cosmology assume that the universe has a  closed geometry, amongst them the Hartle-Hawking no-boundary proposal~\cite{Hartle:1983ai} and the tunnelling proposal~\cite{Vilenkin:1982de,Vilenkin:1984wp}, Whereas the standard interpretation of Coleman-De Luccia vacuum transitions hints at an open universe (however see for instance~\cite{Cespedes:2020xpn} 
for an alternative proposal). For the later stages of the universe, if it is closed ($k=+1$) its spatial volume is finite, whereas if it is flat or open the spatial volume is infinite and therefore the future evolution of the universe would be substantially different.

Despite the fact that the universe appears to be very close to being flat, and that we may never get the precision to be good enough to disentangle its curvature, it is still important to establish to what extent the curvature of the universe may lead to observable implications. If inflation is the correct description of the early universe and the number of e-foldings is much larger than the minimum needed to solve the horizon problem this task may be hopeless. However there is a possibility if the number of e-foldings is close to critical that we may be able to observe signatures of the curvature.

It is a well known fact that curvature perturbations produced during inflation are very close to Gaussian. Deviations from this are called non-Gaussianities and are constrained to be very suppressed by observations~\cite{Planck:2019kim}.  Even so, their detection would reveal crucial information about the physics of the early universe and there are a number of currently running and planned experiments looking for them, including CMB experiments such as CMB-S4~\cite{CMB-S4:2016ple}, The Simons Observatory~\cite{SimonsObservatory:2018koc} and PICO~\cite{Alvarez:2019rhd}; large scale structure surveys including for instance DESI~\cite{Font-Ribera:2013rwa}, SKA~\cite{Maartens:2015mra}, Vera-Rubin telescope~\cite{LSSTScience:2009jmu} and SPHEREX~\cite{Dore:2014cca}.

There has been a large amount of progress in understanding non-Gaussianities from inflation. A particularly important point is that, for a flat universe, single field  attractor models of inflation predict a very weak local non-Gaussianity, which is phrased  as  $f_{\rm NL}^{\rm local} = \frac{5(n_s-1)}{12}$~\cite{Maldacena:2002vr}, with $n_s$ the spectral index. This result shows that measuring a $f_{\rm NL}^{\rm local} = \mathcal{O}(1)$ would mean that at least one of the assumptions behind the derivation of this result is violated, e.g. the single field inflation assumption. Other models of inflation can produce a larger signal  in different limits of the bispectrum shape and so non-Gaussianity becomes a way of distinguishing between different  mechanisms. For instance adding higher derivative interactions can modify the action for the perturbations starting at the quadratic level. This is usually parametrised in terms of the speed of sound $c_s$ at which the perturbations propagate. A low speed of sound can produce a large non-Gaussianity signal $f_{\mathrm{NL}}\sim c_s^{-2}$ which for low $c_s$ can be much larger than the one produced gravitationally~\cite{Chen:2006nt,Cheung:2007st}. Interactions with other fields can also produce different signals, particularly in the squeezed limit~\cite{Sasaki:2006kq,Chen:2009zp,Arkani-Hamed:2015bza}. 
 
Another alternative, which we will explore in this work, is through the presence of curvature in the early universe. The presence of a remnant curvature breaks the scale invariance at large scales producing a particular signal. Since curvature is diluted due to the expansion of the universe this effect is expected to be small, but there may still be room for a possible observational effect, given the fact that current bounds still allow for a detectable effect \cite{Handley:2019tkm,DiValentino:2020hov} (however see \cite{Efstathiou:2020wem}).
 
In this work we develop the formalism to compute the power spectrum of density perturbations for a closed universe and determine the Bunch-Davis vacuum, two-point and higher functions.\footnote{See\cite{Halliwell:1984eu,Seery:2010kh, Ratra:2017ezv,Handley:2019anl,Avis:2019eav}  for related work on closed universes and  \cite{Bucher:1994gb,Yamamoto:1996qq,Sugimura:2013cra} for works on the perturbations on a open Universe} In particular we will compute the non-Gaussian signatures of a closed universe. To do so we will start from a mini-superspace model which can be solved by assuming slow-roll evolution. Using this solution we will build a Schr\"odinger equation for the perturbations which allows us to compute correlation functions of the perturbations at any order. Since the solutions are expressed in terms of spherical harmonics, instead of Fourier modes, they look very different. Nevertheless it is possible to define a flat space limit in which the spherical harmonic modes reduce to Fourier modes. Doing so, we will show that at large scales there are large differences between different geometries. Moreover, since the background evolution is different, if the curvature has not been totally diluted at the time of the CMB, there are interesting observable signals. In particular there is a topological term, i.e. one that depends on whether the Universe is closed or not,  with a three-point function which is not slow-roll suppressed although it is negligible at small scales.

This paper is structured as follows. In Sec.~\ref{sec:Formalism} we consider a minisuperspace model of inflation along with the scalar perturbations for a closed universe. By using the WKB method we show how to solve the equations to find the wave-function  to arbitrary order expanding on $S^3$ harmonics. In Sec.~\ref{sec:PNG} we employ this method to find the primordial non-Gaussian three-point function and we compute a generalisation of $f_{\mathrm{NL}}$ valid on a closed universe. We also study a generalisation of the effective field theory of inflation valid for a closed universe. In Sec.~\ref{sec:Conclusions} we analyse the potential observability of our results and we conclude. 

\paragraph{\textbf{Notation:}}
We will use the global de Sitter metric $ds^2=-dt^2+a(t)^2d\Omega_3^2$ where $t$ is cosmic time, $a(t)$ the scale factor and $d\Omega_3^2$ the metric on $S_2$. This metric covers the whole spacetime and $t$ runs from $-\infty$ in the past to $\infty$ in the future.. We will encounter other notions of time whose notation we summarise here  for easy reference.
\begin{itemize}
\item[$t$]: cosmic Lorentzian time.
\item[$\eta$]: conformal Lorentzian time, running from $-\pi/2$ to $\pi/2$.
\item[$\tau$]: Euclidean time.
\item[$T$]: conformal Euclidean time, running from $-\infty$ to $+\infty$.
\end{itemize}
We define spherical harmonics in $S_3$ as $Y_{p, l m} $ are spherical harmonics labelled by $p$, $l$ and $m$ with $p \geq l \geq 0$ and $-l \geq m \geq l$. 
The Laplacian obeys $\nabla^2_{\mathrm{S}^3}Y_{p, l, m}=-p(p+2)Y_{p, l, m}$, and the functions are normalised such that $\int d\Omega Y_{p, l, m}Y^*_{p', l', m'}=\delta_{pp'}\delta_{ll'}\delta_{mm'}$. Also  $Y_{p, l, -m} = (-1)^{m}Y^*_{p,  l, m}$.

\section{Inflation and interactions in a closed universe}
\label{sec:Formalism}

\subsection{Hamiltonian Approach}

We will start the discussion with a model of  a scalar field coupled to gravity with action given by,
\begin{equation}
S=\int d^4 x\sqrt{-g} \left[\mpl^2 R-\frac{1}{2}\partial_\mu\phi\partial^\mu\phi-V(\phi)\right] \,.
\end{equation}
Assuming that there is a time-dependent homogeneous background with the (FLRW or minisuperspace) metric 
\begin{equation}
ds^{2}=-N^{2}(t)dt^{2}+a^{2}(t)(dr^{2}+\sin^{2}rd\Omega_{2}^{2}) \,. \label{eq:minimetric}
\end{equation}
For convenience we will separate the action for the scale factor $S_g$ from the action for the scalar field $S_m$. In terms of Eq.~\eqref{eq:minimetric} we have that the action becomes $S=S_{g}+S_{m}$, 
with 
\begin{eqnarray}
S_{g} & = &\mpl^2\int dt d^3 x\left(-N^{-1}3a\dot{a}^{2}+3aN\right) \,,\label{eq:Sg}\\
S_{m} & = & \int dt d^3 x\left(N^{-1}\frac{1}{2}a^{3}\dot{\phi}^{2}-Na^{3}((\partial_i\phi)^2+V(\phi))\right) \,.
\end{eqnarray}
Note that the metric assumes a closed universe, with the curvature component appearing in the last term of $S_g$.
We will be interested in potentials that achieve inflation. The equations of motion for the background field $\phi=\phi(t)$ and the scale factor are,
\begin{align}
H^2\equiv\left(\frac{\dot a}{a}\right)^2&=\frac{1}{3\mpl^2}\left(\frac{1}{2}\dot\phi^2+V(\phi)\right)-\frac{1}{a^2} \,,\\
0&=\ddot\phi+3H\dot\phi+V_{,\phi} \,.
\end{align}
Inflation is achieved when the comoving horizon $1/(aH)$ shrinks. This is obtained when $\frac{d}{dt}\left(\frac{1}{aH}\right)<0$, which is equivalent to $\ddot a>0$. This condition can be written as,
\begin{align}
\frac{\ddot a}{a}=H^2\left(1+\frac{\dot H}{H^2}\right) \,,
\end{align}
so inflation lasts as long as $-\dot H/H^2 <1$. This can be phrased in terms of the scalar field using the equations of motion, 
 \begin{align}
 \label{eq:Epsilon}
\frac{\vert\dot H\vert}{H^2}-\frac{1}{a^2H^2}=\frac{1}{\mpl^2}\frac{\dot\phi^2}{2H^2} \equiv \epsilon \,.
 \end{align}
where $\epsilon$ is the slow-roll parameter. Note that in contrast to flat space, there is an extra term that corresponds to the curvature, but both definitions coincide once the curvature dilutes away. Moreover, initially  $\epsilon$ being small does not necessarily imply that the comoving horizon  is shrinking. This puts a  lower bound on the duration of inflation to satisfy current bounds on the curvature.   To avoid confusion  we will  assume that after the curvature dilutes inflation lasts as long as $\epsilon\ll 1$.
In fact at horizon exit for the longest CMB modes the curvature $1/(a^2H^2)\lesssim 0.001$ so  in the region of interest the two definitions are essentially the same.

We will be interested in the perturbations of the scalar field during the slow-roll evolution. This can be obtained by using the ADM formalism and we refer the reader to Appendix~\ref{sec:ADMPert} for a derivation. In terms of the field perturbation $\varphi$ and in flat gauge, the action is given by,

\begin{equation}
S_\varphi=\frac{1}{2}\int dt\ d^3 x\  a^{3}\sqrt{\gamma}\left(N^{-1}\dot\varphi^2-\frac{N}{a^2}\gamma^{ij}\partial_i\varphi\partial_j\varphi+\mathcal{L}_{\mathrm{int}}(N,\varphi)\right) \,,\label{eq:PertAction}
\end{equation}
where $\gamma$ is the metric over the unit 3-sphere and $\mathcal{L}_{\mathrm{int}}(N,\varphi)$ parametrises higher order interactions. As it is shown in Appendix~\ref{sec:ADMPert}, in flat gauge  the dependence on $\mpl^2$ does not appear explicitly. Instead, when the action is  in the comoving gauge the factor of $\mpl^2$ reappears (see Eq.~\eqref{eq:actioncomovinggaguge}).  The action for the perturbations in Eq.~\eqref{eq:PertAction} is then   proportional to $\mpl^2\dot\phi$, which implies that the  perturbation field  only makes sense when the slow-roll parameters are small but non vanishing.

 In order to study the dynamics of this system we will first need the   conjugate momenta,
\begin{align}
&\pi_{N}=0\,, \qquad \pi_{a}=-\mpl^2 N^{-1}6a\dot{a} \,,\label{eq:momenta}\nonumber\\
&\pi_{\phi}=N^{-1}a^{3}\dot{\phi} \,, \qquad \pi_{\varphi}=N^{-1}a^3\sqrt{\gamma}\dot\varphi \,.
\end{align}
The Hamiltonian is $\mathcal{H}=\mathcal{H}_0+\mathcal{H}_\varphi$, ${\cal H}_0$ is the Hamiltonian from the background (minisuperspace) and ${\cal H}_\varphi$ is the Hamiltonian for the perturbations
\begin{align}
{\cal H}_0&=N\left(-\frac{1}{\mpl^2}\frac{\pi_{a}^{2}}{12a}+\frac{\pi_{\phi}^{2}}{2a^{3}}-3a\mpl^2+a^{3}V(\phi)\right) \,,\label{eq:calH}\nonumber\\
{\cal H}_\varphi&=N\left(\frac{1}{\sqrt{\gamma}}\frac{\pi_\varphi^2}{2 a^3}+\sqrt{\gamma}a\gamma^{ij}\partial_i\varphi\partial_j\varphi\right)+{\cal H}_{\mathrm{int}} \,.
\end{align}
We will be interested in solving the Wheeler-de Witt (WDW) equation (the quantum version of the classical Hamiltonian constraint) which, in this case, can be written as $\left(\mathcal{H}_0+\mathcal{H}_\varphi\right)\Psi=0$ and where we need to make the identification,
\begin{equation}
\pi_a=-i\frac{\delta}{\delta a},\qquad\pi_\phi=-i\frac{\delta}{\delta\phi},\qquad\pi_\varphi=-i\frac{\delta}{\delta\varphi}.
\end{equation}

Note that we are neglecting tensor and vector perturbations. For general models  vector perturbations decay during inflation. Tensor perturbations are decoupled up to second order from scalar perturbations. Its action can be studied using similar methods as those that we will describe below.

We will now transform the WDW equation into a Schr\"odinger equation for the perturbations. To do this we will first need to solve the equation for the background. This can be achieved by using the slow-roll conditions.
Indeed, (for a closed universe with $1/(aH)^2\ll 1$) we have that,
\begin{align}
\epsilon= \frac{1}{\mpl^2}\frac{\dot\phi^2}{2H^2}=3 a^2\mpl^2\frac{\pi_\phi^2}{\pi_a^2}\ll 1 \,,
\end{align}
hence in the Hamiltonian $\mathcal{H}_0$ we can neglect the scalar field  momentum since it is negligible compared to $\pi_a$. We solve using the Born-Oppenheimer procedure as used in this context for example in~\cite{Banks:1984cw}. This amounts to expanding in powers of $M_{\rm Pl}^2$ and solving  up to terms $\mathcal{O}(M_{\rm Pl}^0)$. In doing so it is necessary to take the potential to scale as $M_{\rm Pl}^2$ so that in the decoupling limit $M_{\rm Pl}^2\rightarrow\infty$ one has a de Sitter solution for the background. The solution  then takes the form  $\Psi=\psi(a)\chi(a,\varphi)$ where 
$\psi=\frac{1}{\sqrt{dS_0/da}}\exp\left(i\mpl^2 S_0\right)$ corresponds to the  minisuperspace semi-classical wave-function (with pre-factor) and $\chi(a,\varphi)$ corresponds to the perturbations in that background.

$S_0$ is the solution of
\begin{equation}
-\frac{\mpl^2}{12a}\left(\frac{\delta S_0}{\delta a}\right)^2 +a^3 V(\phi)-3a\mpl^2=0 \,,
\end{equation} 
which is the Hamilton-Jacobi equation for minisuperspace. 
Here the function $f(a,\phi)=a^3 V(\phi)-3a\mpl^2$ can be considered as an effective scalar potential for $a$ and $\phi$. For $V>0$ this has a barrier in the $a$ direction for $a<\sqrt{3/V}\mpl$ (remembering that $a$ has the wrong sign of the kinetic term).
In the under the barrier region $a^3V<3a \mpl^2$ we get \footnote{A factor $2\pi^2$ corresponding to the volume of the unit three sphere has been inserted in the formulae for $S_0$.},
\begin{equation}
\label{eq:S0underthebarrier}
S_0=\pm 12\pi^2i \int da a \sqrt{1-\frac{a^2 V}{3\mpl^2}}=\mp i \frac{12\pi^2\mpl^2}{V}\left(1-\frac{a^2V}{3\mpl^2}\right)^{3/2} \,.
\end{equation}
In the over the barrier region $a^3V>3a\mpl^2$ we have instead
\begin{equation}
\label{eq:S0overthebarrier}
S_0=\pm 12\pi^2 \int da a \sqrt{\frac{a^2 V}{3\mpl^2}-1}=\pm  \frac{12\pi^2\mpl^2}{V}\left(\frac{a^2 V}{3\mpl^2}-1\right)^{3/2} \,.
\end{equation}
The two classical solutions imply that $\psi$ is a superposition of two components: under the barrier one has an exponentially rising and an exponentially falling component, while in the classically allowed region the solution is a superposition of expanding and contracting components.
 
\begin{equation}
\pi_a=\mpl^2\frac{\partial S_0}{\partial a}=-12\pi^2\mpl^2N^{-1}a\dot a \,.\label{eq:pi}
\end{equation} 
At $\mathcal{O}(\mpl^0)$ we have a Schr\"odinger equation for $\chi$
\begin{equation}
\frac{i}{12\pi^2 a} \frac{\delta S_0	}{\delta a}\frac{\delta\chi}{\delta a}+\mathcal{H}_\varphi\chi=0 \,, \label{eq:NLO}
\end{equation}
provided that $a$ plays the role of time. In fact from Eq.~\eqref{eq:pi} we see that the first term of this equation is 
$-i\partial /\partial t$ so that we have 
\begin{equation}
\left(i\frac{\partial}{\partial t}-\mathcal{H}_\varphi\right)\chi=0 \,, \label{eq:Schrodinger}
\end{equation}
which corresponds to the Schr\"odinger equation in real time for the modes over the barrier.

For the under the barrier modes there is a factor of $\pm i$ in the expression for $\frac{\partial S_0}{\partial a}$ so that in order to have real $a$ we need to put $t=\pm i \tau$.  Hence Eq.~\eqref{eq:NLO} in the under barrier case becomes
\begin{equation}
\left(\mp\frac{\partial}{\partial \tau}+\mathcal{H}_\varphi\right)\chi=0 \,, \label{eq:EucSchrodinger}
\end{equation}
which is a Euclidean  Schr\"odinger equation for the perturbations. The solutions for $a$ can be obtained by assuming that $\epsilon\ll 1$, which implies that the potential is approximately constant. This of course  reduces to the Euclidean and Lorenzian versions of de Sitter space i.e. \footnote{It is possible to obtain the contracting branch of global de Sitter by a different choice of initial conditions}
\begin{align}
a&= \sqrt{\frac{3\mpl^2}{V}}\sin{\left(\sqrt{\frac{V}{3}}\frac{\tau}{\mpl}\right)},\qquad\mathrm{for}\ a^3V<3a\mpl^2,~0<\sqrt{\frac{V}{3}}\frac{\tau}{\mpl}<\pi/2 \,,\label{eq:a_underthebarrier}\\
a&= \sqrt{\frac{3\mpl^2}{V}}\cosh{\left(\sqrt{\frac{V}{3}}\frac{t}{\mpl}\right)},\qquad\mathrm{for}\ a^3V>3a\mpl^2,\,~t>0 \,.
\label{eq:a_overthebarrier}
\end{align}

 Now the fact that we observe an expanding universe led Vilenkin~\cite{Vilenkin:1986cy} to propose that the wave function in the classical region should have just the outgoing wave component. This is the so-called tunneling wave function. However the WKB matching conditions then imply that one needs both components (the rising as well as the falling) exponentials inside the barrier with equal amplitude (apart from a phase). It appears that this is incompatible with choosing the Bunch-Davies vacuum for the fluctuations. Instead one needs to choose the Hartle-Hawking (HH) wave function~\cite{Hartle:1983ai}\footnote{For a recent discussion see \cite{deAlwis:2018sec, DiTucci:2019bui}  where it is shown that the tunneling wave function while it has Gaussian fluctuations as is the case for the HH wave-function, is unstable to quartic fluctuations unlike the HH one.}.

 In the classically allowed region the latter is a cosine function, implying that it is a superposition of expanding and contracting universes. Thus we need to assume that the two components de-cohere and so we may just focus on the expanding one where the Schr\"odinger equation in Eq.~\eqref{eq:Schrodinger} is satisfied. On the other hand in the classically forbidden region one has only the rising component (corresponding to the lower sign in Eq.~\eqref{eq:S0underthebarrier}) which is effectively exponentially suppressed at $a=0$.  We note therefore for future reference that going from the real time Schr\"odinger equation in Eq.~\eqref{eq:Schrodinger} to the Euclidean version in Eq.~\eqref{eq:EucSchrodinger} (with the lower sign) corresponds to putting $t=+i\tau$.
 
 \subsubsection{Bunch-Davies vacuum}

To find the vacuum let us first solve the equations of motion for the quadratic actions for the perturbation. We assume that spacetime is a fixed de Sitter such that the metric is given by \begin{align}
a^2(t)=\frac{1}{H^2}\cosh^2 (H t) \,,
\end{align}
where $t$ is cosmic time and we defined $\mpl^{2}H^2=V/3$.
It is convenient to decompose $\varphi$ in terms of spherical harmonics
\begin{equation}
\varphi(x)=\sum _{p, l, m}\varphi_{p, l, m}(t)Y_{p, l, m}(\Omega) \,,
\end{equation}
where $Y_{p, l m} $ are spherical harmonics labelled by $p$, $l$ and $m$ with $p \geq l \geq 0$ and $-l \geq m \geq l$. 
The Laplacian obeys $\nabla^2_{\mathrm{S}^3}Y_{p, l, m}=-p(p+2)Y_{p, l, m}$, and the functions are normalised such that $\int d\Omega Y_{p, l, m}Y^*_{p', l', m'}=\delta_{pp'}\delta_{ll'}\delta_{mm'}$. Also  $Y_{p, l, -m} = (-1)^{m}Y^*_{p,  l, m}$.

In this section we will find the solution which corresponds to the Bunch-Davies vacuum both in the Euclidean region (i.e. under the barrier) and in the Lorentzian region (i.e. over the barrier).

\subsubsection*{Euclidean region}

Let us first solve the region under the barrier.
Due to the spherical symmetry the Euclidean action becomes, 
\begin{equation}
\label{eq:EuclideanQuadraticAction}
S_2=\sum_{p, l, m}\int d\tau \ a^3\left(|\varphi'_{p, l, m}|^2+p(p+2)\frac{|\varphi_{p, l, m}|^2}{a^2}\right) \,.
\end{equation} 
Where $'$ denotes derivative with respect to Euclidean time $\tau$.
The Hamiltonian density is then
\begin{align}
H=\sum_{p, l, m} \bigg[\frac{\pi_\varphi^2}{a^3}+p(p+2)a \varphi_{p, l, m} \bigg] \,.
\end{align}
Let us assume the Gaussian ansatz for the wave-function 
\begin{equation}
\label{eq:chi}
\chi=\exp\bigg(\sum_{p, l, m}c_p(\tau)\varphi_{p, l, m}\varphi_{p, l, -m}\bigg) \,,
\end{equation}
so the time dependence is in the coefficient $c_p$ whereas $\varphi_p$ is a field profile, and where $\varphi_{p, l, m}^*=\varphi_{p, l, -m}$. By making the identification
\begin{align}
\pi_\varphi\to-i\frac{\delta}{\delta\varphi} \,,
\end{align}
in the Hamiltonian of Eq.~\eqref{eq:EucSchrodinger} then gives, using the ansatz in Eq.~\eqref{eq:chi},
\begin{align}
\partial_\tau c_p=-a^{-3}c_p^{2}+a p(p+2) \,.
\end{align}
It is also convenient to write the equations in term of conformal Euclidean time such that, $ds^2=a(T)^2(dT^2+d\Omega_3^2)$ where $T$ goes from $-\infty$ in the past to $\infty$ in the future. Then the scale factor in Eq.~\eqref{eq:a_underthebarrier} becomes,
\begin{equation}
\label{eq:EuclideanScaleFactor}
a(T)=H^{-1}\sech(T) \,,
\end{equation}
and the Schr\"odinger equation in Eq.~\eqref{eq:Schrodinger} becomes
\begin{equation}
\partial_T c_p +a^{-2} c_p^2-a^{2}p(p+2)=0 \,.
\end{equation}
To solve this equation we put $c_p=a^2\partial_T f_p/f_p$ where $f_p$ satisfies the linear eqn.
\begin{equation}
\partial_T\left(a^2 \partial_T f\right)=a^2(p(p+2))f \,.
\end{equation}
This is in fact the ‘radial’ equation coming from the 4D Laplace equation in the metric of Eq.~\eqref{eq:minimetric} in Euclidean conformal coordinates in the gauge $N=1$, i.e. $ds^2=a^2(T)(dT^2+d\Omega_3^2)$. To solve this equation we take  the scale factor  given by the de Sitter solution  $a(T)=H^{-1}\sech(T)$. Imposing the solutions to be regular at $a=0$ which in these coordinates correspond to $T\to-\infty$, we get that the mode function is,\footnote{In the case the field has a mass $m_\phi$ the equation  changes to $\partial_T\left(a^2 \partial_T f\right)=a^2(p(p+2))f+ a^4 m_\phi^2 f$ the solutions are
\begin{align}
f_p(T)&=C_1 \left(\frac{\sech(T)}{2}\right)^{p/2} \, _2F_1\left(p+\Delta,p+3-\Delta;p-1;\frac{1}{2} -\frac{\tanh (T)}{2}\right)\nonumber\\&+C_2 \left(\frac{\sech(T)}{2}\right)^{p/2} \, _2F_1\left(p+\Delta,p+3-\Delta;p+2; \frac{\tanh (T)}{2}-\frac{1}{2}\right) \,,
\end{align}
where $\Delta=3/2-\nu=3/2-\sqrt{\frac{9}{4}-\frac{m_\phi^2}{H^2}}$.
\begin{align}
f_p(T)=C_1(\sinh(T)-(p+1)\cosh(T))e^{(p+1)T} \,.
\end{align}
The normalization factor $C_1$ can be found by imposing the canonical commutation relations (we will show it explicitly in the Lorentzian case below).}
\begin{equation}
\label{eq:modefunctionEuclid}
f_p(T) = \frac{H (\sinh(T)-(p+1)\cosh(T))e^{(p+1)T}}{\sqrt{2p(p+1)(p+2)}} \,.
\end{equation}
For future reference, given Eq.~\eqref{eq:modefunctionEuclid}, one can compute the functions $c_p = a^2 \partial_T f_p/f_p$:
\begin{equation}
\label{eq:clE}
c_p = \frac{1}{H^2} \frac{p (p+2)\sech^2 T}{\cosh^2 T (1 + p - \tanh T)} \,.
\end{equation}

\subsubsection*{Lorentzian region}

In the over the barrier region the wave-function is given by $\chi \simeq\exp(i\sum_{p, l, m}c_p(\eta)\varphi_{p, l, m}\varphi_{p, l, -m})$. The scale factor is now given by Eq.~\eqref{eq:a_overthebarrier}, which in conformal time $\eta$ is given by
\begin{equation}
\label{eq:LorentzianScaleFactor}
a(\eta)=H^{-1}\sec(\eta) \,,
\end{equation}
where $\eta$ runs from $0$ to $\pi/2$ which corresponds to the classical  expanding phase that starts  at $\eta=0$. The equation for the perturbations is,
\begin{equation}
-\partial_\eta(a^2\partial_\eta f_p)=a^2p(p+2)f_p \,, \label{eq:Lorentizanequation}
\end{equation}
where  $c_p=- a^2\partial_\eta f/f$. We find the solution to this equation in Appendix~\ref{sec:solution_eq} to be  given by
\begin{equation}
f_p=C_1((1+p)\cos\ \eta+i\sin\ \eta )e^{-i(p+1)\eta}+C_2 ((1+p)\cos\ \eta-i\sin\ \eta )e^{+i(p+1)\eta} \,.
\label{eq:generalsolutionLorentzian}
\end{equation}
In order to define a vacuum let us recall that in flat de Sitter this is defined by picking an incoming wave  deep inside the bulk. In this limit the physical momenta $k/aH\gg 1$ and the spacetime can be well approximated by Minkowski spacetime. In other words the Bunch-Davies vacuum is defined as the one that reduces to the Minkowski vacuum in the sub-horizon limit. For a closed universe we cannot take arbitrarily small values of $a$ without going under the barrier. A solution that leads to a Bunch-Davies vacuum  is to impose that the mode function in Eq.~\eqref{eq:generalsolutionLorentzian} has to be also a solution in the Euclidean region~\cite{Laflamme:1987mx}. From Eq.~\eqref{eq:modefunctionEuclid} we find that this can only happens when $C_2=0$ in Eq.~\eqref{eq:generalsolutionLorentzian} since the Euclidean limit correspond to $\eta\to iT$ the choice of sign being that corresponding to the Hartle-Hawking choice for the wave function $\psi$ - see paragraph below Eq.~\eqref{eq:a_underthebarrier}.\\

In order to fix the remaining constant we impose the Klein-Gordon normalization over the mode function.
This reduces to $\vert C_1\vert^2(2p(p+1)(p+2))=H^2$ so we finally find that 
\begin{equation}
f_p=H \frac{((p+1)\cos\ \eta+i\sin\ \eta )e^{-i(p+1)\eta}}{\sqrt{2p(p+1)(p+2)}} \,. \label{eq:modefunction}
\end{equation}
We would like to compare this result to the flat space mode function $f_k\sim\frac{e^{-ik\eta}}{k^{3/2}}(i+k\eta)$ where $\eta$ is conformal time and $a=-1/(H\eta)$. The analogy  is not exact but if we look at the limit $\eta\to\pi/2$ then both scale similarly at large $l$: $f_k \sim k^{-3/2}$ and $f_p \sim p^{-3/2}$. Moreover when $\eta$ is close to zero the closed mode function scales similarly as the Minkowski vacuum. For future reference we also write down the expression for the functions $c_p = - a^2 \partial_\eta f_p/f_p$:
\begin{equation}
\label{eq:clL}
c_p = i \frac{p (p+2) \sec^2 \eta}{H^2 (1 + p + i \tan \eta)} \,.
\end{equation}
Let us now make some comments about these results. Notice that in getting Eq.~\eqref{eq:modefunction} we need to impose conditions over Eq.~\eqref{eq:generalsolutionLorentzian} at $a=0$. In the more general case that the condition is imposed at other value of the scale factor it will imply that there are growing and decaying modes at $\dot a=0$ instead of only decaying. This then translates into a more general solution for the coefficients of Eq.~\eqref{eq:generalsolutionLorentzian}.\\

Before continuing, let us stress again  that  the  derivation of the mode functions assumed that under the barrier the scalar field was almost static. This assumption can in principle  be dropped provided that there is a  classical solution. This is because, in the end to form a Schr\"odinger equation all that we need is to use a classical solution to introduce a clock. Generalisations to other potentials have been considered  in~\cite{Banks:1984np,Halliwell:1984eu}. The fact that there it is possible to define a WKB expansion in the context of slow-roll  inflation has  explained more recently in~\cite{Janssen:2020pii}.

\subsubsection*{Wave function and two-point correlators}

Let us now compute  the wave-function.\footnote{Alternatively it is possible to define the fields in the Schr\"odinger picture~\cite{Luscher:1985iu,Long:1996wf} where the fundamental object is the wave-function coefficient rather than the mode function.} Plugging back Eq.~\eqref{eq:clL} into the wave-function we have that,
 \begin{align}
 \label{eq:WaveFunction1}
 \chi=\exp\left(-i\sum_{p, l, m}   \frac{\sec \eta}{2H^2} \frac{p (p+2)}{i(p+l)\cos \eta-\sin \eta}\varphi_{p,l, m}\varphi_{p, l, -m}\right)\,.
 \end{align}
Expanding for $\eta$ close to $\pi/2$ we find
 \begin{align}
 \chi=\exp\left(i\sum_{p,l,m}\frac{1}{2H^2}\left(-\frac{p(p+2)}{\eta-\pi/2}+i p(p+1)(p+2)+\mathcal O(\eta-\pi/2)\right)\varphi_{p, l, m}\varphi_{p, l, -m}\right) \,.
 \label{eq:wavefunction_latetime}
 \end{align}
Notice that the divergent piece is a pure phase while the leading order quadratic contribution is given by $\chi \simeq \exp(-\sum_{p,l,m}p(p+1)(p+2)\vert\varphi_{p, l, m}\vert^2)$. Notice that picking the other Euclidean solution (which corresponds to the analytic continuation $T\to-i\eta$) the real piece picks an overall plus sign and the perturbations grow. 

From the wave-function we can compute the two point function to be,
 \begin{align}
 \langle\varphi_p^2\rangle= \frac{H^2}{2p(p+1)(p+2)} \,.
 \label{eq:2point_function}
 \end{align}
 In order to compare it to observations we need to compute the curvature perturbation $\zeta$, which at linear order  and after horizon crossing is given by $\zeta_*=-(H/\dot\phi) \varphi_*$ where the asterisk denotes horizon crossing.\footnote{This corresponds to transforming to the  comoving gauge which is  explained in Appendix~\ref{sec:ADMPert}.} Then we have
 \be
\boxed{ \langle \zeta_p\rangle^2=\frac{H^4}{\dot\phi^2}\frac{1}{2p(p+1)(p+2)}=\frac{H^4}{2\mpl^2 (\dot H-\frac{1}{a^2})}\frac{1}{2p(p+1)(p+2)}}
 \label{eq:2point_function2}
\ee
which for large $p$ leads to the flat space power spectrum (which scales as $H^2/(\epsilon  k^{3})$).

Both, the wave function in Eq.~\eqref{eq:wavefunction_latetime} and the two point function generalise the results obtained for a flat universe in~\cite{Maldacena:2002vr}.
In general to compute the correlation functions we use the relation
 \begin{equation}
 \langle \chi\vert \mathcal{O}[\varphi,\pi]\vert\chi\rangle=\int\mathcal{D}\varphi\ \chi^*[\varphi]\mathcal{O}\left[\varphi,-i\frac{\delta}{\delta\varphi}\right]\chi[\varphi]\label{eq:pathintegral}
\end{equation}
where we are assuming that the wavefunciton $\chi$ is implicitly normalised by a factor $\sqrt {N}$. Using this relation we find
\begin{align}
\langle\varphi_{p, l, m}\varphi_{p, l, -m}\rangle&=N\int\mathcal{D}\varphi |\varphi_{p, l, m}|^2e^{i {(S-S^*)}}\nonumber\\
&=N\int\mathcal{D}\varphi |\varphi_{p, l, m}|^2 e^{\sum_{p, l, m}-2\mathrm{Im}{c_p}|\varphi_{p, l, m}|^2}=\frac{1}{2\mathrm{Im}{c_p}} \,.
\end{align}
Note that this does not depend on $l, m$ so we can write
 \begin{align}
 \langle\varphi_p\varphi_{p}\rangle'=\frac{1}{2\text{Im}\  c_p}=H^2\frac{(1+p)^2\cos^2\eta+\sin^2\eta}{2p(p+1)(p+2)} \,,
 \label{eq:2pointfunction}
 \end{align}
 which implies that at $\eta=\pi/2$ becomes, $\langle\varphi_p^2\rangle\sim 1/(p(p+1)(p+2))$ as we saw before. The scale dependence at large scales can have an effect on the large angular distances of the CMB, which will be similar to the one described in~\cite{Cespedes:2020xpn}. This will depend on the duration of inflation as well as  whether there are higher derivative interactions that change the two point function. We will discuss a modified speed of sound effect in Sec.~\ref{sec:EFT}.

Generalising Eq.~\eqref{eq:chi} to include terms beyond the quadratic order, we can write $\chi=\exp\left(i\Gamma\right)$ and 
\begin{equation}
\label{eq:Gamma}
\Gamma=\sum_{p, l,m} c_p(\tau)\varphi_{p, l, m}\varphi_{p, l, -m}+\frac{1}{6} \sum_{p_i, l_i, m_i}G_{123} c^{(3)}_{p_1,p_2,p_3}\varphi_{p_1, l_1, m_1} \varphi_{p_2, l_2, m_2} \varphi_{p_3, l_3, m_3}+... \,,
\end{equation}
where $G_{123} =\int d\Omega \, Y_{p_1, l_1, m_1} Y_{p_2, l_2, m_2}Y_{p_3, l_3, m_3}$ (this can be expressed in terms of 3j symbols).

One can compute higher order correlation functions following the same procedure that we used for the power spectrum. The three-point function and four-point function coefficients are given by

 \begin{align}
 \label{eq:3PointCF}
 &\langle\varphi_{p_1, l_1, m_1}\varphi_{p_2, l_2, m_2}\varphi_{p_3, l_3, m_3}\rangle'=\frac{2\text{Im}c^{(3)}_{p_1,p_2,p_3}}{\prod_i 2 \text{Im} c_{p_i}} \,,
 \end{align}
 \begin{footnotesize}
 \begin{align}
  \label{eq:4PointCF}
  &\langle\varphi_{p_1, l_1, m_1}\varphi_{p_2, l_2, m_2}\varphi_{p_3, l_3, m_3}\varphi_{p_4, l_4, m_4}\rangle'=\frac{1}{8}\left(\prod_{i=1}^{4}\frac{1}{\text{Im} c_{p_i}}\right)\left( \text{Im}c^{(4)}_{p_1,p_2,p_3,p_4}-\sum_{j=i}^{4}\frac{\text{Im}c^{(3)}_{p_1,p_2,p_j}\text{Im}c^{(3)}_{p_3,p_4,p_j}}{\text{Im}c_{p_j}}-\mathrm{perms.}\right) \,, \nonumber
 \end{align}
 \end{footnotesize}
where the prime denotes geometric factors.\\

Note that in order for the wave-function to be normalisable all the higher-order interactions should be regarded as small corrections to the Gaussian wave-function.

\subsection{Interactions}

Adding interactions is straightforward in the Schr\"odinger picture (see for example ~\cite{Cespedes:2020xqq} for a recent implementation in the case of a flat universe). In order to compute the higher order coefficients it is better to write the Hamiltonian as
 \begin{equation}
\mathcal{H}[\varphi, \pi] =\mathcal{ H}_{\rm{free}}[\varphi, \pi] + \mathcal{H}_{\rm{int}}[\varphi, \pi] \,,
\end{equation}
where $\mathcal{H}_{\rm{free}}$ is the free Hamiltonian from Eq.~\eqref{eq:calH}. From now on we will drop the $l$ and $m$ indices from the scalar field.

Let us start assuming that we have a cubic interaction $H_{\rm{int}}=a^3 \lambda \varphi^3$. Using Eq.~\eqref{eq:Schrodinger} with $\Gamma$ given in Eq.~\eqref{eq:Gamma} and taking  functional derivatives we get up to third order in the fields~\cite{Cespedes:2020xqq}
\begin{equation}
\left(-\partial_\eta+3 a^{-2}\sum_{i=1}^3c_{p_i}(\tau) \right)c^{(3)}_{p_1,p_2,p_3}=a \frac{\delta^3 \mathcal{H}_\mathrm{int}}{\delta\varphi_{p_1}\delta\varphi_{p_2}\delta\varphi_{p_3}} \,,\label{eq:3HJ}
\end{equation}
where $\mathcal{H}_\mathrm{int}$ is also re-written in term of spherical harmonics such that the coefficient $G_{123}$ factors out from the equation.
It is best to transform into the interaction picture performing an unitary transformation at each time to define a new operator, $\phi_p^I=\phi_p/f_p$. This corresponds to rescaling  the wave-function coefficients,
\begin{equation}
\label{eq:InteractionPictureCoefficients}
c^I_{p_1,p_2,p_3}=f_{p_1}f_{p_2}f_{p_3}c^{(3)}_{p_1,p_2,p_3} \,.
\end{equation}
The interaction picture Hamiltonian is defined as $\mathcal{H}^{I}_{\rm{int}}[\phi^I]=\mathcal{H}_{\rm{int}}[f_p\phi_l^p]$. This greatly simplifies the equations, for example for a cubic interaction as in Eq.~\eqref{eq:3HJ} it becomes,
\begin{equation}
-\partial_\eta c^I_{p_1,p_2,p_3}=a\frac{\delta^3H_\mathrm{int}^I}{\delta\varphi_{p_1}^I\delta\varphi_{p_2}^I\delta\varphi_{p_3}^I}=a^4 \lambda  f_{p_1}f_{p_2}f_{p_3} \,,
\label{eq:cubicHJ}
\end{equation}
where we have used that $c_p=-a^2\partial_\eta f_p/f_p$. The resulting expression is simpler to integrate. Indeed we have that the interaction coefficient is given by

\begin{align} c^I_{p_1,p_2,p_3}&=\lambda \int_{-\infty}^\eta  d\eta'\  a^4 f_{p_1}f_{p_2}f_{p_3}\\
 &=\frac{\lambda}{H^4}\int_{-\infty}^{\eta} d\eta' \sec{\eta'}^4\prod_{i=1}^3 \frac{((1+p_i)\cos\ \eta'+i\sin\ \eta' )e^{-i(p_i+1)\eta'}}{\sqrt{2p_i(p_i+1)(p_i+2)}} \,.
\end{align}
Note that  $\eta<0$ corresponds to the Euclidean region  which we get by tha continuation $\eta\to i T$. In this region the mode functions are given by Eq.~\eqref{eq:modefunctionEuclid} and the scale factor by Eq.~ \eqref{eq:EuclideanScaleFactor}. In the end, since the mode function has a factor of $\exp(i(p(p+1)T))$, the integrand goes to zero  in the limit $T\to\infty$. This is similar to the contour rotation done in the in-in computations on a flat universe~\cite{Maldacena:2002vr}, where the integrals vanish at $a=0$. 
After performing the integral we find,
\begin{align}
c^{(3)}_{p_1,p_2,p_3}&=\frac{\lambda}{H^4}\prod_{p=1}^3\frac{1}{(p_i+1)\cos\eta+i\sin\eta}\nonumber\\
  & \left(-\frac{i\sec (\eta )}{3} \sum_{\mathrm{perm}}^3\left(\frac{\sec^2\eta}{3}-i (p_1+1)\tan\eta+p_1^2-p_1 p_2-1\right)\right. \nonumber\\
   &\left.+\frac{2 i e^{i \eta }\sum_i p_i (p_i+1)
   (p_i+2)}{3 (p_T+2) } \, _2F_1\left(1,\frac{1}{2}
   (-p_T-2);\frac{1}{2} (-p_T);-e^{2
   i \eta }\right)\right) \,,
\end{align}
where $p_T = p_1 + p_2 + p_3$. Note that the piece that contains the hypergeometric function diverges at $\eta=\pi/2$, so this expression needs to be regularised before taking the limit. To see this let us expand around $\eta_0$ with $\pi/2-\eta_0\ll1$. We get
\begin{align}
c^{(3)}_{p_1,p_2,p_3}&=\frac{\lambda}{H^4}\left(\frac{1}{9} \left(\frac{24}{(\pi -2 \eta_0 )^3}+\frac{\sum_{i=1}^3 (9 p_i (p_i+2)+4)}{\pi -2 \eta_0 }\right)\right.\nonumber\\
&+\left.\frac{i}{9} (p_T+3) \left(6 p_T+2+\sum_{\rm{perm}} p_1(4 p_1-(p_2+p_3))\right)\right.\nonumber\\
&\left. - \frac{i}{3} \, \sum_i p_i (p_i+1)
   (p_i+2)\left(\gamma_E+\frac{1}{3}+i\pi/2+\log(\pi-2\eta_0)+\psi ^{(0)}\left(-\frac{p_T}{2}-1\right)\right)\right) \,,
\end{align}
where $\psi^0$ is the digamma function, $\gamma_E$ the Euler-Mascheroni constant and   we have been careful in first taking the limit in $c_{p_1,p_2,p_3}^I$ and then divide by the mode functions evaluated at $\eta=\pi/2$. Notice that  $\eta_0$ acts as a cutoff up to where the correlation function is valid.
The fact that there are IR divergencies in de Sitter correlation functions is a known fact~\cite{Weinberg:2006ac,Burgess:2009bs,Seery:2010kh,Burgess:2010dd,Giddings:2011ze,Anninos:2014lwa, Cespedes:2020xqq}. It has been studied extensively for a flat universe in the past so it is not surprising that it also appears on closed universes.  

Now let us compute the correlation function close to the conformal boundary. First notice that the first two pieces are real so they don't contribute to the correlation function. Indeed we have,
\begin{align}
\langle\varphi_{p_1}\varphi_{p_2}\varphi_{p_3}\rangle'&=\frac{2\lambda}{H^4}\prod\frac{1}{2p_i(p_i+1)(p_i+2)}\left( \frac{2(p_T+3)}{9} \left(6p_T+2+\sum_{\rm{perm}}p_1(4p_1-(p_2+p_3))\right)\right.\nonumber\\
&\left.+\frac{2}{3} \,\sum_i p_i (p_i+1)
   (p_i+2)\left(\gamma_E+\log(\pi-2\eta_0)+\psi ^{(0)}\left(-\frac{p_T}{2}-1\right)\right)\right) \,,
&
\label{eq:coeffphi3}
\end{align} 
which is similar to the expression computed for a flat universe~\cite{Creminelli:2011mw},
\begin{align}
\langle\varphi_{k_1}\varphi_{k_2}\varphi_{k_3}\rangle'&\propto \frac{1}{\prod_i k_i^3}\left(\sum_ik_i^3(-1+\gamma_E+\log(-k_T\eta_*))+k_1k_2k_3-\sum_{i\neq j}k_i^2k_j\right) \,,
\end{align}
where now  $\eta_*$ is evaluated in the Poincaré patch coordinates and so is close to zero. Furthermore notice that for large $p_T$ one has that  $\psi_0(-1-p_T)\sim\log (-p_T)$. This suggests that both  expressions coincide for large $l$ which can be shown explicitly by assuming that $p_1,p_2,p_3\gg 1$.


\section{Primordial non-Gaussianity}
\label{sec:PNG}


\subsection{Non-Gaussianities in a flat universe}

We will be interested in computing correlation functions  of the curvature perturbation $\zeta$. In this section we will collect some useful results in flat space. First, the two and three point functions are usually written as,
\begin{align}
\langle\zeta(\k_1)\zeta(\k_2)\rangle&=(2\pi)^3\delta(\k_1+\k_2)P(k_1) \,, \label{eq:PowerSpectrumFlat}\\
\langle\zeta(\k_1)\zeta(\k_2)\zeta(\k_3)\rangle&=(2\pi)^3\delta\left(\sum \k_i\right)B(k_1,k_2,k_3) \label{eq:BispectrumFlat}\,,
\end{align}
where $P(k)$ is the power spectrum and  $B(k_1,k_2,k_3)$ the bispectrum. The power spectrum is also written in terms of the scale invariant power spectrum through
\begin{align}
\Delta_\zeta^2=\frac{k^3}{2\pi^2}P(k) \,.
\end{align}
The value of $\Delta_\zeta$ can be obtained from  CMB observations  and is  $\Delta_\zeta^2= 2.09\times 10^{-9}$~\cite{Planck:2018jri}.
Non-Gaussianity is usually parametrised in terms of $f_{\mathrm{NL}}$ defined as,
\begin{align}
B(k_1,k_2,k_3)=\frac{18}{5}\frac{f_{\mathrm{NL}}(k_1,k_2,k_3)}{(k_1 k_2 k_3)^2}\Delta_\zeta^4 \,.
\end{align}
Bounds on $f_\mathrm{{NL}}$ depends on  the relative size of the wavenumbers. Since  they form a triangle, $f_\mathrm{{NL}}$  is usually constrained depending on the shapes they form. For instance, when all three $k$ are equal, the shape is called equilateral. Non-Gaussianity is also constrained from CMB observations, for instance the equilateral shape is bounded to be  $f_{\mathrm{NL}}^{\mathrm{equil}}=-26\pm 47$ whereas for the orthogonal shape $f_{\mathrm{NL}}^{\mathrm{ortho}}=-38\pm 24$~\cite{Planck:2019kim}. Notice that since the bispectrum is proportional to $\Delta_\zeta^4$ non-Gaussianity is extremely suppressed.

\subsection{Non-Gaussianities in a closed universe}

After understanding how to compute higher order correlation functions, let us now obtain the bispectrum due to primordial non-Gaussianities. In flat gauge the action at cubic order is given\footnote{To the authors' knowledge this action was obtained first in~\cite{Clunan:2009ib}. There is an effective mass term there i.e. $\partial^2\rightarrow \partial^2+3$ which we ignore since in the regime of interest $3\ll p(p+2)$.} by,
\begin{align}
S_3=\frac{1}{\mpl}\int dt d^3 x \sqrt{\gamma} \left(\frac{a}{\sqrt{2\epsilon}}\varphi^3+a^5\sqrt{2\epsilon} H\dot\varphi^2 \partial^{-2}\dot\varphi +\frac{\delta \mathcal{L}_2}{\delta\varphi}f(\varphi)\right) \,,
\label{eq:cubic_action}
\end{align}
with $\epsilon = -\frac{\dot\phi^2}{2H^2}$ and where $f(\varphi)$ is  given by,
\begin{align}
f(\varphi)= \frac{\sqrt{2\epsilon}}{8}\varphi^2-\frac{\sqrt{2\epsilon}}{4}\partial^{-2}(\varphi\partial^{2}\varphi)-\frac{3\sqrt{2\epsilon}}{8}(\partial^{-2})\varphi^2 \,.
\label{eq:def_fieldrd}
\end{align}
The terms proportional to the equations of motion can be removed by a field redefinition, 
\begin{align}
\label{eq:FieldRedefinition}
\varphi=\varphi_c+f(\varphi_c) \,.
\end{align}
In term of $\varphi_c$ the action becomes,
\begin{align}
S_3=\frac{1}{\mpl}\int d^4x \left(\frac{a}{\sqrt{2\epsilon}}\varphi_c^3+a^5\sqrt{2\epsilon} H\dot\varphi_c^2(\delta)^{-2}\dot\varphi_c \right) \,.
\end{align}
The first term only appears when there is curvature. Its appearance comes from the fact that $V'''$ has a factor of $a^{-2}$ due to the curvature term in the Friedmann equation,
\begin{equation}
V'''(\phi)=\frac{6H}{a^2\dot{\phi}}+O(\sqrt{\epsilon}) \,. \nonumber
\end{equation}
Moreover, notice that this term is not slow-roll suppressed and it appears only on a closed (or open) universe and it dilutes with the expansion of the universe. 
Including all the terms the Hamiltonian is given by,
\begin{align}
\mathcal{H}_{\rm{int}}=-\frac{1}{\mpl}\int  d^3 x\left(\frac{a^2}{\sqrt{2\epsilon}}\varphi_c^3+\frac{\sqrt{2\epsilon}H}{a^3}\pi_\varphi^2\partial^{-2}\pi_\varphi\right) \,.
\label{eq:IntHamPNG}
\end{align}
Expanding the fields into spherical harmonics, the Hamilton-Jacobi equation is the following\footnote{We write explicitly the expressions in the Lorentzian region, using the conformal time $\eta$. Analogous expressions can be found in the Euclidean regions, using conformal Euclidean time $T$.},
\begin{align}
\left(\partial_\eta-\frac{1}{a^2}\sum_i c_{p_i}\right)c^{(3)}_{p_1,p_2,p_3}=\frac{1}{\mpl}\frac{a^2}{\sqrt{2\epsilon}}+\frac{1}{\mpl}\frac{\sqrt{2\epsilon}H}{a^3}\sum _{\rm{perm}}c_{p_1}c_{p_2}\partial^{-2}c_{p_3} \,.
\end{align}
Since we will only compute the cubic coeffcient $c^{(3)}$ from now on we will remove the superindex on the coefficient and we will only keep the subindices.
Written in terms of the interaction picture this equation simplifies to,
\begin{align}
\partial_{\eta}c^I_{p_1,p_2,p_3}=\frac{1}{\mpl}\frac{a^2}{\sqrt{2\epsilon}}\prod_{i}f_{p_i}+\frac{1}{\mpl}\frac{\sqrt{2\epsilon} H}{a^3}\prod_{i}f_{p_i}\sum _{\rm{perm}}c_{p_1}c_{p_2}\partial^{-2}c_{p_3} \,,
\label{eq:IntegraPNG}
\end{align}
where the expressions for the functions $f_p$ and $c_p$ are given in Eq.~\eqref{eq:modefunction} and Eq.~\eqref{eq:clL} respectively. In Eq.~\eqref{eq:IntegraPNG}, the second term is the standard contribution that also appears in the spatially flat Universe case. The second term however, is specific to a spatially closed geometry.\\

We can compute third order coefficient applying the  the machinery developed in Sec.~\ref{sec:Formalism}. We will call the first term in  Eq.~\eqref{eq:IntegraPNG} $c^{a}$, whereas the second term will be denoted $c^{b}$: the result of this will have to match with the flat case result in an appropriate limit that will be discussed below. Replacing the Laplacian operator by its eigenvalue $-p(p+2)$, the third order coefficient is given by,
\begin{align}
c^{(b)}_{p_1,p_2,p_3}&=\frac{1}{\mpl}\frac{\sqrt{2\epsilon_*} H_*}{\prod_{i}f_{p_i}}\int_{-\infty}^{\pi/2} d\eta \, a^{-3}\prod_{i}f_{p_i}\sum_{\rm{perm}}c_{p_1}c_{p_2}\frac{1}{-p_3(p_3+2)}c_{p_3}\,,\nonumber
\end{align}
where, in computing the integral, in the $-\infty < T < 0$ region we have used the Euclidean expressions for $a$, $f_p$, $c_p$ (see Eq.~\eqref{eq:EuclideanScaleFactor}, Eq.~\eqref{eq:modefunctionEuclid} and Eq.~\eqref{eq:clE}), while we have used Lorentzian expressions (see Eq.~\eqref{eq:LorentzianScaleFactor}, Eq.~\eqref{eq:modefunction} and Eq.~\eqref{eq:clL}) in the region $0 < \eta < \pi/2$. Note that the contribution to the integral from the Euclidean region vanishes, hence evaluating the result at $\eta = \pi/2$ we are left with,
\begin{equation}
c^{(b)}_{p_1,p_2,p_3} = i\frac{1}{\mpl}\frac{\sqrt{2\epsilon_*}}{H_*^2} \,  \times \frac{\sum_{i>j}p_i(p_i+2)p_j(p_j+2)}{p_T+3} \,,
\end{equation}
where the asterisk denote the value at horizon crossing. This is equivalent to the flat space case, where the contour is rotated in the past, $\eta\to-(1+i\epsilon)\infty$ to select the interacting vacuum. Notice that the cubic coefficient has an overall $i$, which implies that there it is a small contribution in the wave function since it appears schematically as $\chi\sim\exp(- l^3 \varphi^3)$. Finally, using Eq.~\eqref{eq:3PointCF}, the three point function is given by 
\be
\boxed{\langle\varphi_{c, p_1}\varphi_{c, p_2}\varphi_{c, p_3}\rangle'=\frac{\sqrt{2\epsilon_*}}{4}\frac{H_*^4}{\mpl}\prod_{i}\frac{1}{p_i(p_i+1)(p_i+2)}\times\frac{\sum_{i>j}p_i(p_i+1)p_j(p_j+1)}{p_T+3}} \,.
\ee
where the $'$ denotes that we have to multiply the right hand side by $G_{123}$.
At this point we can compute the coefficients from the field redefinition in Eq.~\eqref{eq:FieldRedefinition}. In order to do so, we need to use Wick's theorem: for instance the first term in the field redefinition gives the following contribution to the three point function,
\begin{align}
\langle \varphi(x_1)\varphi(x_2)\varphi(x_3)\rangle=\langle \varphi_c(x_1)\varphi_c(x_2)\varphi_c(x_3)\rangle+\frac{\sqrt{2\epsilon}}{4}\left(\langle\varphi(x_1)\varphi(x_2)\rangle\langle\varphi(x_1)\varphi(x_3)\rangle+\mathrm{perm.}\right) \,.
\end{align}
The first two terms in Eq.~\eqref{eq:def_fieldrd} also appear on a flat universe. For a closed universe their contribution is~\cite{Clunan:2009ib},
\begin{align}
\label{eq:3pfFR12}
\langle  \varphi_{p_1}\varphi_{p_2}\varphi_{p_3}\rangle_{\rm{fr}}'=\frac{\sqrt{2\epsilon_*}}{32}\frac{H_*^4}{\mpl}\prod_{i}\frac{1}{p_i(p_i+1)(p_i+2)}\times\left(\sum_i p_i(p_i+1)(p_i+2)+\sum_{i\neq j}(p_i+1)p_j(p_j+2)\right)
\end{align}
where  the subscript denotes that are terms from the field redefinition.
Finally, the last term from the field redefinition gives,
\begin{align}
\label{eq:3pfFR13}
\langle \varphi_{p_1}\varphi_{p_2}\varphi_{p_3}\rangle_{\rm{fr}}'=\frac{3\sqrt{2\epsilon}}{32}\frac{H_*^4}{\mpl}\prod_{i}\frac{p_T+3}{p_i(p_i+1)(p_i+2)} \,.
\end{align}
Summing all the terms, the three point function is given by,
\begin{small}
\begin{align}
\langle \varphi_{p_1}\varphi_{p_2}\varphi_{p_3}\rangle'&=\frac{H_*^4}{\mpl}\frac{\sqrt{2\epsilon_*}}{32\prod_{i}p_i(p_i+1)(p_i+2)}\\
&\times\left(\sum_i p_i(p_i+1)(p_i+2)+\sum_{i\neq j}(p_i+1)p_j(p_j+2)+8\frac{\sum_{i>j}p_i(p_i+2)p_j(p_j+2)}{p_T+3}+3(p_T+3)\right) \,. \nonumber
\end{align}
\end{small}
In order to compare with observations we  need to change to  $\zeta$ gauge. The relation between the two gauges simplifies after horizon crossing and it can be written using~\cite{Sasaki:1995aw},
\begin{align}
\zeta (t,\x)= \sum_{n=1}^{\infty}\frac{1}{n!}\left(\frac{\partial ^nN(t,t^*)}{\partial\phi_*^n}\right)(\varphi_*(t_*,x))^n \,,
\end{align}
where $t$ is some time after horizon crossing $t^*$,  and  $N$ is the number of efolds between the spatial slices labelled by $t_*$ and $t$. Note that this formula is written in real space and the transformation to spherical harmonics  has to be treated carefully \cite{Clunan:2009ib}.  In the end this means that there  are extra terms in the  bispectrum of $\zeta$. Indeed we have,
\begin{align}
\langle \zeta_{p_1}\zeta_{p_2}\zeta_{p_3}\rangle'=\left(\frac{\delta N}{\delta\phi_*}\right)^3\langle\varphi_{p_1}\varphi_{p_2}\varphi_{p_3}\rangle+\frac{1}{2}\left(\frac{\delta N}{\delta\phi_*}\right)^2 \frac{\delta^2 N}{\delta \phi_*^2}\left(\sum_{\rm{perm}}\langle \varphi_{p_1}\varphi_{p_1}\rangle\langle \varphi_{p_2}\varphi_{p_2}\rangle\right) \,.
\label{eq:3pfZetaCoV}
\end{align}
Given that $\delta N=H dt$ we finally have  that,
\begin{align}
\frac{\delta N}{\delta \phi_*}=-\frac{H_*}{\dot\phi_*} \,,\qquad \frac{\delta^2 N}{\delta\phi_*^2}=\frac{H_*\ddot \phi_*}{\dot\phi^3_*}+\frac{1}{2\mpl^2}-\frac{1}{\dot\phi_*^2 a^2} \,,
\end{align} with all of this we finally have,
\begin{small}
\begin{align}
\label{eq:closedBispectrum2V}
\langle \zeta_{p_1}\zeta_{p_2}\zeta_{p_3}\rangle'&=\frac{H_*^4}{\mpl^4}\frac{1}{2\epsilon_*\times 32}\frac{1}{\prod_{i}p_i(p_i+1)(p_i+2)}\\
&\times\left(\sum_i p_i(p_i+1)(p_i+2)+\sum_{i\neq j}(p_i+1)p_j(p_j+2)+8\frac{\sum_{i>j}p_i(p_i+2)p_j(p_j+2)}{p_T+3}+3(p_T+3)\right) \nonumber\\
&+\frac{H_*^4}{\mpl^4}\frac{1}{4\epsilon}\left(\frac{1}{2\epsilon}\left(\frac{\ddot \phi_*}{\dot\phi_*H_*}+\frac{1}{a_*^2 H_*^2}\right)-\frac{1}{2}\right)\frac{1}{\prod_i 2p_i(p_i+1)(p_i+2)}\sum_i p_i(p_i+1)(p_i+2) \,,\nonumber
\end{align}
\end{small}
where in the last line we have used Eq.~\eqref{eq:2point_function}. 
This result is a generalization of Maldacena's computation~\cite{Maldacena:2002vr}, 
\begin{align}
\langle\zeta_{\k_1}\zeta_{\k_2}\zeta_{\k_3}\rangle'=\frac{H^4_*}{\mpl^4}\frac{1}{4\epsilon^2}\frac{1}{\prod_i 2k_i^3}\left[\epsilon\left(\sum_i k_i^3+\sum_{i\neq j}k_ik_j^2+\frac{8}{k_T}\sum_{i>j}k_i^2 k_j^2\right)+\frac{2\ddot\phi_*}{\dot\phi_* H_*}\sum_i k_i^3\right] \,,
\label{eq:PNG_flatU}
\end{align}
where a prime here means that we are taking out a factor of $(2\pi)^3\delta(\sum_i k_i)$. Our result for a closed Universe agrees with the result in Eq.~\eqref{eq:PNG_flatU} for a flat Universe in the limit of large $p$. 

However, as we will see below, the most important difference between the flat Universe and the closed Universe cases is determined by the first term in Eq.~\eqref{eq:IntegraPNG}, that we have neglected so far. In fact, it gives,
\begin{align}
c^{(a)}_{p_1,p_2,p_3}&=\frac{1}{\sqrt{2\epsilon\mpl} \prod_i f_{p_i}}\int^{\pi/2}_{-\infty} d\eta \, a^2\prod_i f_{p_i}\label{eq:ca}\\
&\simeq-\frac{i}{\sqrt{2\epsilon}\mpl H^2}\left(-(p_T+1)+\frac{p_1p_2 p_3}{(p_T+2)(p_T+4)} + \frac{\sum_{i>j} p_i p_j}{(p_T+2)}\right) - \frac{1}{\sqrt{2\epsilon} H^2}\frac{1}{\eta-\pi/2} \,, \nonumber
\end{align}
where in the last equation we have expanded in powers of $\eta-\pi/2$. As before, in the Euclidean region $(-\infty, 0)$ we have used Eq.~\eqref{eq:EuclideanScaleFactor}, Eq.~\eqref{eq:modefunctionEuclid} and Eq.~\eqref{eq:clE}, and the contribution to the integral from this region vanishes.
Note that there is no logarithmic divergence in this case, and moreover that the infrared divergence does not appear in the correlation function because the term proportional to $(\eta - \pi/2)^{-1}$ is real. The difference with Eq.~\eqref{eq:coeffphi3} arises because the $\varphi^3$ term in Eq.~\eqref{eq:IntHamPNG} has an extra $a^2$ which appears after integrating by parts the slow-roll coefficients. This changes the degree of the divergence in such a way that the correlation function is now finite. 
Finally, using Eq.~\eqref{eq:3PointCF}, we have that the contribution to the bispectrum by the new terms is given by, 
\begin{align}
\langle \varphi_{p_1}\varphi_{p_2}\varphi_{p_3}\rangle'=\frac{H_*^4}{4\sqrt{2\epsilon_*}\mpl}\frac{1}{\prod_{i}p_i(p_i+2)(p_i+2)}\left(-(p_T+1)+\frac{p_1p_2p_3}{(p_T+2)(p_T+4)}+\frac{\sum_{i>j}p_i p_j}{(p_T+2)}\right) \,.
\label{PNG}
\end{align}
Writing this expression in the comoving gauge we find, 
\begin{align}
\langle \zeta_{p_1}\zeta_{p_2}\zeta_{p_3}\rangle'=\frac{H_*^4}{\mpl^2}\frac{1}{4(2\epsilon)^2}\frac{1}{\prod_{i}p_i(p_i+2)(p_i+2)}\left(-(p_T+1)+\frac{p_1 p_2 p_3}{(p_T+2)(p_T+4)}+\frac{\sum_{i>j}p_i p_j}{(p_T+2)}\right) +\dots\,.
\label{eq:zeta3prime}
\end{align}
where the dots are terms that depend on the second piece of Eq.~\eqref{eq:3pfZetaCoV} which are already included in Eq.~ \eqref{eq:closedBispectrum2V}.
Notice that the bispectrum  we have  computed is $1/\epsilon$ larger than the result in Eq.~\eqref{eq:closedBispectrum2V}. In any case for small scales this contribution vanishes since it  scales as $p^{-8}$ whereas the terms in Eq.~\eqref{eq:closedBispectrum2V} scale as $p^{-6}$. Please note that our expressions in Eq.~\eqref{eq:closedBispectrum2V} and Eq.~\eqref{eq:zeta3prime} are valid for any $p$ and they agree with~\cite{Seery:2010kh} at very large $p$.

Before continuing let us comment on the soft theorems in the presence of  curvature. First let us notice  that when one of the modes is much larger than the others $k_3\ll k_1,k_2$, it means that it exits the horizon earlier. Its overall effect over shorter modes is to make them cross the horizon earlier by $\delta t\sim -\zeta/H$. For a flat universe this lead to a consistency condition~\cite{Maldacena:2002vr,Creminelli:2004yq} which is related to the existence of a soft mode. When there is curvature this consistency condition is broken, as it has been pointed out in~\cite{Avis:2019eav}, due to the fact that the curvature breaks the cosmological soft theorems.

\subsection{Computing $f_\mathrm{{NL}}$}

As mentioned before, the bispectrum for a flat universe is usually written as~\cite{Baumann:2014nda},
\begin{align}
B_\zeta(k_1,k_2,k_3)=\frac{18}{5}\frac{f_{\mathrm{NL}}(k_1,k_2,k_3)}{(k_1k_2k_3)^2}\Delta_\zeta^4 \,, \label{eq:flatFnl}
\end{align}
where $f_\mathrm{{NL}}$  depends on the three momenta\footnote{Notice that the definition used in~\cite{Maldacena:2002vr,Clunan:2009ib} is \begin{align}
B_\zeta(k_1,k_2,k_3)=\frac{12}{5}f_{\mathrm{NL}}\frac{\sum k_i^3}{\prod 2 k_i^3}\Delta_\zeta^4 \,.
\end{align} } and $\Delta_\zeta^2=k^3/2\pi^2 \langle\zeta_\k^2\rangle$. This definition removes the overall scale dependence and so $f_\mathrm{{NL}}$ depends only on the ratios between the three comoving momenta. Usually the ratios are then fixed by considering  different shapes, for instance when the momenta form an equilateral triangle. 

In order to generalise this parametrisation let us first note that the definition of $\Delta_\zeta^2$ contains a geometric factor that  does not appear on a closed universe. Instead, let us use the definition
\begin{align}
\tilde{\Delta}_\zeta^2\equiv \frac{H_*^4}{\mpl^2}\frac{1}{4\epsilon}
\end{align}
such that $\tilde{\Delta}_\zeta^2=p(p+1)(p+2)\langle\zeta_p^2\rangle$.
Next, let us notice that there is a factor of $(2\pi)^3\delta(\sum \vec{k_i})$ from the definition of the bispectrum in Eq.~\eqref{eq:BispectrumFlat}. This corresponds to a geometric factor, and for a closed universe it is possible to see from 
\eqref{eq:relation_delta_G}, that it matches the sum of $3js$ functions for large $l$ and small $r$. We can define the bispectrum on a flat universe as,
\begin{align}
\langle\zeta_{p_1}\zeta_{p_2}\zeta_{p_3}\rangle'=B_\zeta(p_1,p_2,p_3)
\end{align} then in analogy with the flat case  we can define  the Non Gaussian parameter $\tilde f_{\mathrm{NL}}$ through the relation,
\begin{align}
B_{\zeta}(p_1,p_2,p_3)=\frac{18}{5}\frac{\tilde f_{\mathrm{NL}}(p_1,p_2,p_3)}{\left(\prod_i p_i(p_i+1)(p_i+2)\right)^{2/3}}\tilde\Delta_\zeta^4 \,, \label{eq:closedFnl}
\end{align}
which reduces to Eq.~\eqref{eq:flatFnl} for large $p$ up to a geometric proportionality constant.
Notice that, since the bispectrum is not scale invariant $\tilde f_{\mathrm{NL}}$ will also depend on the sizes of $p$, not only on their ratios. Nevertheless, as we show in Sec.~\ref{sec:Harmonics}, in the limit when $p\gg 1$ the modes are well approximated by their flat universe counterpart. In this case  $\tilde f_{\mathrm{NL}}$ reduces to the flat space  $f_{\rm NL}$. 

In this sense $\tilde f_{\mathrm{NL}}$ incorporates the scale dependent effects that are relevant at small $p$. From the bispectrum in Eq.~\eqref{eq:closedBispectrum2V} we can see that the scale dependent  contribution scales as $\epsilon^2 p^{-2}$, which is the reason the bispectrum is suppressed at small scales. On the other hand the contribution from Eq.~\eqref{PNG} is not slow-roll suppressed and hence it becomes the dominant term at small $p$.
 
In order to estimate the size of non Gaussianites let us compute the bispectrum in the equilateral limit when all $p's$ have  the same size. We can write the result schematically as, 
 \begin{align}
\boxed{
 \tilde{f}_{\mathrm{NL}}=f_{\mathrm{NL}} - \frac{5}{12} \frac{1}{a^2 H^2} +\frac{85}{162}\frac{1}{p^2}}
 \end{align}
where $f_{\mathrm{NL}}\sim\mathcal{O}(\epsilon,\eta)$ is the usual non-Gaussianity computed on a flat universe. Corrections from the curvature to those terms will be typically of order $\sim\mathcal{O}(\epsilon,\eta)(1-p^{-2}+\mathcal{O}(p^{-3}))$ and so very small. At large scales there is a larger contribution from the term proportional to $3(p_T+3)$ which scales as $p^{-2}$ but it is slow-roll suppressed. More interesting are the other terms. The second one corresponds to the remnant curvature and it is bounded by Planck to be very suppressed. The last term corresponds the the first term in Eq.~\eqref{eq:IntHamPNG}, which as we discussed appears because the third derivative of the potential has a leading order piece that depends on the curvature. Depending on the size of the slow-roll parameters, the last term can be the dominant one by several orders of magnitude. This illustrates the fact that curvature effects may eventually have observational implications.

Let us comment on which modes are observable. We can estimate that the largest CMB scale  given the comoving distance to the CMB dipole is around at $\chi_L=.46$. This corresponds to an angular fraction of $p\sim\pi/.46\sim 7$ ~\cite{Seery:2010kh}. Hence the smallest $p$ observed in the sky is $p\sim 7$, by which the largest contribution to $f_{\mathrm{NL}}$ is at most of order $.01$. Nevertheless, notice that assuming only Planck data a closed universe is preferred with $\Omega_k\sim -0.056\pm 0.01$~\cite{Planck:2018jri,Handley:2019tkm}, such a value will imply a larger set of available modes and hence the maximum value of  $f_{\mathrm{NL}}$ increases by an order of magnitude.
\subsection{Effective field theory of inflation}
\label{sec:EFT}
With our formalism it is also possible to include other interactions that appear, for example,  when including higher derivative interactions such as in $P(X)$ theories. In order to do so it is better to use the  the EFT of inflation \cite{Cheung:2007st,Creminelli:2013cga}. The idea is to consider the most general action with broken time diffeomorphisms expanded around an FLRW background,
\begin{align}
S=\int &d^4 x \sqrt{-g}\left[\frac{1}{2}\mpl^2R+\mpl^2\left(\dot H-\frac{k}{a^2}\right)\cdot g^{00}-\mpl^2\left(3H^2+\dot H+2\frac{k^2}{a^2}\right)\right.\nonumber\\
&\left.+\frac{1}{2!}M(t)(g^{00}+1)^2+\frac{1}{3!}c_3(t)M(t)^4(g^{00}+1)^3+...\right] \,.
\end{align}
The Goldstone boson $\pi$ associated to broken time translations is introduced through the St\"uckelberg trick, $t\to t+\pi$. Up to cubic order, and  neglecting the mixing with the graviton, this leads to the following action
\begin{align}
S=\int d^4 x\sqrt{-g} \, \frac{\epsilon H^2\mpl^2}{c_s^2}\left[\left(\dot\pi^2-c_s^2\frac{(\nabla\pi^2)}{a^2}\right)+\alpha\dot\pi\frac{(\nabla\pi)^2}{a^2}+\beta\dot\pi^3\right] \,,
\label{eq:EFTOI}
\end{align}
where $g$ is the minisuperspace metric determinant and the speed of sound $c_s$ is defined as,
\begin{align}
c_s^{-2}=1-\frac{2M^4}{\mpl^2(\dot H-ka^{-2})} \,,
\label{eq:cs}
\end{align}
and,
\begin{align}
\alpha=c_s^2-1\,,\qquad \beta=(1-c_s^2)\left(1+\frac{2}{3}\frac{c_3}{c_s^2}\right) \,.
\end{align}
There is an extra term in the quadratic action proportional to $\dot H\pi^2$ which we have ignored since it is subleading order in slow roll. Notice however, that this term  will induce an additional  curvature term  which will act as an effective mass term. As before, we will ignore it since in the regime of interest $3\ll p(p+2)$.
 At cubic order the action will contain in Eq.~\eqref{eq:EFTOI} contain an additional  curvature term, similar to that described in Eq.~\eqref{eq:cubic_action}, since it is not slow-roll suppressed. Its effect will be similar to the one described in the previous section. Even though the parameters of the action look identical to the ones on the flat universe case they are not the same. The difference is due to the fact that they are obtained after expanding around different backgrounds, as  has been pointed out in~\cite{Creminelli:2013cga}. When comparing to the flat universe EFT, it can be shown that the parameters for a closed universe  can depend on higher order coefficients. Notice the explicit dependence on $k$ on Eq.~\eqref{eq:cs}.

In order to compute observables we need to relate the  Goldstone mode to the curvature perturbation. This is done by a  gauge transformation, from unitary gauge to comoving gauge, which on super-horizon scales becomes $\zeta=-H\pi$. Finally, since the action is decoupled from gravity it is possible to  formulate it the problem in terms of a Schr\"odinger equation by assuming that it corresponds to a scalar field $\pi$ coupled to a fixed gravitational background. This is very similar to the action of the perturbations we obtained through the Born-Oppenheimer procedure, although in this case we will not specify the minisuperspace action. 

In order to write the Hamiltonian it is convenient to define the gravitational coupling scale $f_\pi^4\equiv 2\mpl^2 \vert\dot H-1/a^2\vert c_s^{-2}$  . In terms of this the  conjugate momenta is,
\begin{align}
P_\pi\equiv \frac{\partial\mathcal L}{\partial\dot\pi}=\sqrt{-g} f_\pi^4\dot\pi \left(1+\alpha \frac{(\nabla \pi)^2}{a^2}+3\beta\dot\pi\right) \,.
\end{align} 
Then the Hamiltonian density is,
\begin{align}
\mathcal{H}=\frac{1}{\sqrt{-g}}\frac{1}{2f_\pi^4}P_\pi^2+\sqrt{-g}\frac{f_\pi^4c_s^2}{2a^2 }\nabla\pi^2-\frac{\alpha}{2a^2}P_\pi(\nabla\pi)^2+\frac{1}{g}\frac{\beta}{2f_\pi^8}P_\pi^3 \,.
\end{align}
Let us first solve the free part of the Schr\"odinger equation. Expanding the fields in spherical harmonics we have that 
\begin{align}
\mathcal{H}_0=\sum_{p, l, m}\frac{1}{2f_\pi^4}\frac{P_\pi}{a^3}+c_s^2a p(p+2)\pi_{p, l, m}^2 \,.
\end{align}
In order to solve the Schr\"odinger equation, again we assume a Gaussian wave-function
\begin{equation}
\chi=\exp(i\sum_{p, l, m} c_p(\eta)\pi_{p, l, m} \pi_{p, l, -m}) \,,
\end{equation}
and we make the replacement
$P_\pi\to i\frac{\delta}{\delta_\pi}$. Doing so we obtain the Hamilton-Jacobi equation
\begin{align}
\frac{\partial c_p}{\partial\eta}=\frac{1}{a^2f_\pi^4} c_p^2+c_s^2a^2 p (p+2) \,,
\end{align}
which can be  solved by assuming a Bunch-Davies vacuum. We find that the solution is,
\begin{align}
c_p= \frac{f_\pi^4 c_s^2}{H^2} \frac{ p(p+2)\sec\eta}{i\sqrt{1+c_s^2 p(p+2)}\cos\eta - \sin\eta} \,.
\label{eq:cl_EFT}
\end{align}
We are interested in the two point function of $\zeta$ at $t\to\pi/2$ which is  
\begin{align}
\langle\zeta^2\rangle&=H^2\langle\pi^2\rangle=\frac{H^4}{2c_s^2 f_\pi^4}\frac{1}{p(p+2)\sqrt{1+c_s^2 p (p+2)}}\nonumber\\
&=
\frac{H^4}{4 \mpl^2(\dot H-1/a^2)}\frac{1}{p(p+2)\sqrt{1+c_s^2 p(p+2)}} \,.
\end{align}
This result generalises Eq.~\eqref{eq:2point_function2} to the case with a modified speed of sound. In order to compare with a flat universe we have that for large $p$ the power spectrum is given by,
\begin{align}
\langle\zeta^2\rangle= \frac{H^4}{4\mpl^2(\dot H-1/a^2)c_s}\frac{1}{p^3}\left(1-\frac{3}{p}+\frac{15 c_s^2-1}{2c_s^2}\frac{1}{p^2}+\mathcal{O}\left(p^{-3}\right)\right) \,.
\end{align}  notice that in for large $p$ we recover the usual  enhancement of  a factor of $c_s^{-1}$. We can also see that there is a further enhancement which appears  for low $p$ modes. Notice that in the case of small $c_s$ the second term can be of order $c_s^{-1}$ changing the large scale power spectrum. This behaviour  has been highlighted before~\cite{Avis:2019eav} but using an approximated version of the mode functions. 
\subsubsection*{Non-Gaussianities}
Let us now compute the three point function from the EFT. Since we are interested in the scaling behaviour it will be enough  to compute the $\dot\pi^3$ This can be done in a similar fashion to how we computed non-Gaussianities in Sec.~\ref{sec:PNG}, but with the difference that the wave-function coefficients are different. Let us start by computing the  cubic coefficient for $\mathcal{H}_{\rm{int}}=\frac{\beta}{2f_\pi^4}\frac{\pi_\varphi^3}{a^6}$. 
The Hamilton-Jacobi equation for the cubic coefficient is,
\begin{align}
\left(\partial_\eta -\frac{1}{a^2}\sum_i c_{p_i}\right)c_{p_1,p_2,p_3}=\frac{\beta }{2 f_\pi^4a^5}\prod_{i=1}^3 c_{p_i} \,,
\end{align}
 where $c_p$ is given by Eq.~\eqref{eq:cl_EFT}. As usual it is easier to  compute the cubic coefficient in the interaction picture,
\begin{align}
c^I_{p_1,p_2,p_3}&= \frac{\beta }{2 f_\pi^8} \int_{-\infty}^{\eta} d\eta'\  a^{-5}\prod_{i=1}^3 f_{p_i}(\eta')c_{p_i}(\eta') \,,
\end{align}
where in this case the mode function is given by,
\begin{equation}
f_p\sim(\sqrt{1+c_s^2 p(p+2)}\cos(\eta)+i\sin(\eta))e^{-i\sqrt{1+c_s^2 p(p+2)}\eta} \,,
\end{equation}
and $f_\pi^2a^2\partial_\eta(\log f_p)=c_p$.
Performing the integral as we have done in Sec.~\ref{sec:PNG} and using Eq.~\eqref{eq:InteractionPictureCoefficients}, we find that the cubic coefficient evaluated at $\eta = \pi/2$ is given by,
\begin{align}
c_{p_1,p_2,p_3}=\frac{if_\pi^4  c_s^6 \beta}{H}\frac{p_1(2+p_1)p_2(2+p_2)p_3(2+p_3)}{(-2+\sum_i\sqrt{1+c_s^2 p_i(p_i+2)})(\sum_i\sqrt{1+c_s^2 p_i(p_i+2)})(2+\sum_i\sqrt{1+c_s^2 p_i(p_i+2)})} \,,
\end{align}
where we have used that in the $\eta\leq 0$ the mode function is given by its analytic continuation $\eta\to-iT$ such that the integral converges  at $a=0$. The cubic coefficient has an additional dependence on the speed. At leading order in the small parameter $1-c_s \ll 1$ we find,
\begin{equation}
c_{p_1,p_2,p_3} \simeq \frac{i f_\pi^4 \beta}{H}\frac{2 p_1 (p_1+2) p_2 (p_2+2) p_3
   (p_3+2)}{(p_T+1) (p_T+3)
   (p_T+5)} \,.
\end{equation}
Notice that the result coincides with one obtained in~\cite{Cespedes:2020xpn} for large $p$. Nevertheless the structure is different from flat space where correlation functions contain single poles. These poles have been related to the flat space limit of the correlation functions~\cite{Maldacena:2011nz,Raju:2012zr}. In this case we see that the poles are displaced due to the curvature. It will be interesting to investigate this issue further. Moreover it seems that for large curvature there are no single poles.

\begin{figure}[h!]
  \includegraphics[scale=0.55,trim=2.1cm 10cm 1cm 4cm,clip]{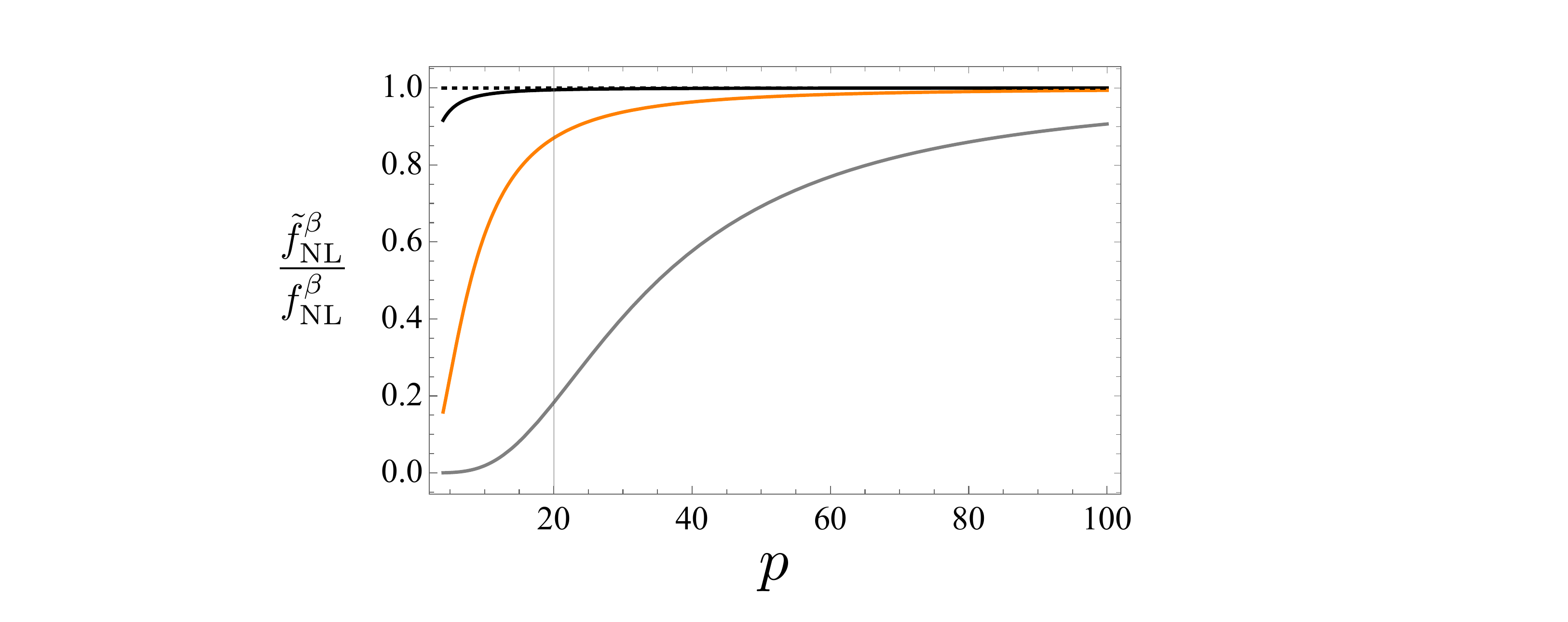}
\caption{We plot the ratio between the curved space $\tilde {f}_{\mathrm{NL}}$ and the flat space $ f_{\mathrm{NL}}$. The black line corresponds to $c_s=0.9$, the orange line to $c_s=0.2$. and the grey line to $c_s=0.05$. We see that the effect of the curvature is enhanced when $c_s$ decreases. This is due to the fact that a lower speed of sound lowers the scale at which modes cross the horizon. Since the impact of the curvature is larger on large wavelength modes the the three point function is greatly modified.   \\}
\label{Fig:fnl_cs}
\end{figure}
Using Eq.~\eqref{eq:3PointCF} and Eq.~\eqref{eq:Epsilon} we find the three point function,
\begin{align}
\langle\zeta_{p_1}\zeta_{p_2}\zeta_{p_3}\rangle&=\frac{\beta c_s^4 H^4}{8 \mpl^4\epsilon^2}\prod_{i=1}^3\frac{1}{\sqrt{1+c_s^2 p_i(p_i+2)}}\\
&\times\frac{1}{(-2+\sum_i\sqrt{1+c_s^2 p_i(p_1+2)})(\sum_i\sqrt{1+c_s^2 p_i(p_1+2)})(2+\sum_i\sqrt{1+c_s^2 p_i(p_1+2)})} \,. \nonumber
\end{align}

Notice that this expression scales as $\Delta_\zeta^4\beta$ for large $l$, which coincides with the flat space result~\cite{Senatore:2009gt}. Nevertheless for $1\ll p\lesssim 1/\sqrt{\epsilon}$ the scaling is different. For instance for the equilateral shape, we have that $\tilde f_{\rm NL}$ behaves as
\begin{align}
\tilde{f}^{\beta}_{\mathrm{NL}}=\beta\frac{9c_s^2 p^2(p+1)^2(p+2)^2}{(1+c_s^2 p(p+2))^2(5+9 c_s^2 p(2+p))}\sim \frac{1}{c_s^2}\left(1+\left(1-\frac{23}{9c_s^2}\right)\frac{1}{p^2}+\mathcal{O}(p^{-3})\right) \,.
\end{align}
Let us unpack this formula. We first have that there is an  overall scaling $c_s^{-2}$, which as has been pointed out before~\cite{Chen:2006nt,Cheung:2007st} can imply a larger signal than the gravitational coupling we have discussed. Moreover for small $c_s$ there is an extra contribution at small $p$ which suppresses the value of $\tilde{f}_{\mathrm{NL}}$.  This is a feature particular to the  curvature of the universe,  notice that the overall scaling is $c_s{-n}p^{-n}$  which implies that  at $c_s$ decreases other terms in the expansion also become relevant. The suppression depends heavily on $c_s$, as it can be seen from the plot in  Fig. \ref{Fig:fnl_cs}, for $c_s\sim 1$ the signal $\tilde{f}_{\mathrm{NL}}$ is  suppressed at $p \sim \mathcal{O}(10)$ similarly than for the gravitational couplings. More interesting a small speed of sound causes an important effect which last up to $p\sim 100$.  Note that the speed of sound has to satisfy the bound $c_s\gg 0.003$ in order the EFT to not become strongly coupled. In any case even for mild values of $c_s$ there is a large effect of the curvature on primordial non-Gaussianity.


\section{Conclusions}
\label{sec:Conclusions}

In this paper we have computed correlation functions of density perturbations generated during inflation in a closed universe. Even though it is well known that the inflationary expansion dilutes the curvature, it is still an important question to ask whether it leads to any signature.

The most important part of the paper is the developing of the formalism to compute the density perturbations in a closed form. In Sec.~\ref{sec:Formalism} we were able to solve the equations determining the Schr\"odinger wave-function for a closed universe in a closed form. This differs from the flat universe case, as in a closed universe the wave-function is determined in terms of spherical harmonics relevant for the $S^3$ symmetries rather than the Fourier expansion which is typical of flat case. Yet we were able to compare with the flat space limit and agreement was found. The importance of having a closed form is that our results hold for any value of the modes $p$. In Sec.~\ref{sec:Formalism} we determined the corresponding Bunch-Davis vacuum, the two-point function and interactions after computing the wave-function in the Schr\"odinger formalism. In general we find that curvature effects may play a relevant role specially at large scales for values of $p\lesssim \mathcal{O}(10)$, even though in terms of observations, this regime is limited due to cosmic variance.

In Sec.~\ref{sec:PNG} we concentrated on non-Gaussianities, generalising the result of Maldacena to closed universes, illustrating the differences at large scales and checking that the two results agree at small scales as they should. Computing $f_{\mathrm{NL}}$ allowed us to explicitly see how its expression includes explicit dependence on the curvature.
We have shown that the cubic order contains a term which is larger than the usual slow roll non-Gaussianity, and might lead to interesting observable consequence. 

Some of the results obtained here were discussed previously in the literature \cite{Seery:2010kh}. Nevertheless most of them had to take the approximation of small curvature (and large $p$) at the cost of not being able to accurately compute the largest contribution from the curvature. We checked that our results reduce to the results of~\cite{Maldacena:2002vr,Seery:2010kh} for large values of $p$ as expected. Furthermore, using the effective field theory of inflation we computed the effects on non-Gaussianities of a low speed of sound as it happens in higher derivative theories.

Overall, the curvature effects on non-Gaussianities are small but not totally negligible. We found terms that contribute to non-Gaussianities that may lead to observable consequences even if a long period of inflation dilutes the curvature. It may be interesting to explore these potential observational implications in more detail. One of the most promising tests of non-Gaussianity is the scale dependence of the halo bias. Since this an effect important over large scales, one could expect that our results here will lead to an enhanced signal. Still more work is needed to disentagle the scale dependent signal from the curvature, but importantly our results show that for gravitational interactions the leading non-Gaussian  signal comes from the curvature, which further suggest the importance of understanding the imprints of the curvature on large scale structure observables. 

The techniques used in this article may be generalised in several ways, if the universe is an expanding bubble, the boundary effects may be relevant, furthermore extending our results to the open universe case should be doable, noting that the underlying symmetries are different and the nature and mode expansion of the wave-function will require a separate calculation.
 It will also be possible to study interactions with other particles. The effect of the curvature at large scales suggest that there might be important effects on the cosmological collider~\cite{Arkani-Hamed:2015bza} which will be interesting to explore.  

\section*{Acknowledgements}
We  thank conversations with Santiago Avila and Will Handley. We especially thank Paolo Creminelli and  Enrico Pajer for carefully reading a preliminary version of this manuscript. The work of FQ has been partially supported by STFC consolidated grants ST/P000681/1, ST/T000694/1. FM is funded by a
UKRI/EPSRC Stephen Hawking fellowship, grant reference EP/T017279/1 and partially supported by the
STFC consolidated grant ST/P000681/1. The work of SC has been funded by a Contrato de Atraccion de Talento (Modalidad 1) de la Comunidad de Madrid (Spain), number 2017-T1/TIC-5520 and the IFT Centro de Excelencia Severo Ochoa Grant SEV-2016-0597.

\appendix
\section{Derivation of  the action by solving the Hamiltonian constraints}
\label{sec:ADMPert}
In this section we detail how to obtain $S_2$.
Let us consider the following ADM decomposition,
\begin{equation}
ds^2=-N^2dt^2+h_{ij}\left(dx^i+N^idt\right)\left(dx^j+N^jdt\right) \,,
\end{equation}
where $h_{ij}$ is the perturbed metric on spatial slices. For scalar perturbations can be written as,
\begin{equation}
h_{ij}=a(\tau)^2e^{2\zeta}\gamma{ij} \,,
\end{equation}
where $\zeta$ is the curvature perturbation and $\gamma$ is the three metric on the spehere. The action becomes,
\begin{equation}
S=\frac{1}{2}\int \ d t\ d^3x\left[N\left(R^{(3)}-2V-h^{ij}\partial_i\phi\partial_j\phi\right)+\frac{1}{N}\left(E_{ij}E^{ij}-E^2+(\dot\phi-N^i\partial_i\phi)^2\right)\right] \,.
\end{equation}
The equations for the shift and the lapse become the momentum and Hamiltonian constraints,
\begin{eqnarray}
\nabla_i\left(N^{-1}(E^i_j-\delta^i_j E)\right)-\frac{1}{N}\left(\dot\phi-N^j\partial_j\phi\right)\partial_i\phi=0 \,,\\
R^{(3)}-2V-N^{-2}(E_{ij}E^{ij}-E^2)-N^{-2}(\dot\phi-N^i\partial_i\phi)^2-h^{ij}\partial_i\phi\partial_j\phi=0 \,.
\end{eqnarray}
We fix to the flat gauge by chosing
\begin{equation}
\delta\phi=\varphi \,, \hspace{1cm} h_{ij}=a^2\Omega_{ij} \,.
\end{equation}
To solve the constrain equations we expand
\begin{equation}
N=1+\delta N,\hspace{1cm} N_i=\partial_i\psi \,.
\end{equation}
At leading order,
\begin{eqnarray}
\Delta N&=&H\left(\frac{1}{2}\dot\phi\varphi+\psi\right) \,,\\
\psi&=&\epsilon(\triangle+3-\epsilon)^{-1}\left(\frac{d}{dt}\left(-\frac{H}{\dot\phi}\varphi\right)+\frac{a^2}{\dot\phi}\varphi\right) \,,
\end{eqnarray}  
where $\triangle$ is the Laplacian on the unit three sphere. Using the constraints in the action and integrating by parts we get
\begin{equation}
S_2=\frac{1}{2}\int dt\ d^3 x\  a^{3}\sqrt{\gamma}\left(\dot\varphi^2-\frac{1}{a^2}\gamma^{ij}\partial_i\varphi\partial_j\varphi\right) \,.
\end{equation}
In order to compare to observations a better suited function is the curvature fluctuation $\zeta$
that appears in comoving gauge,
\begin{align}
\delta\phi=0 \,, \quad h_{ij}=a^2e^{2\zeta}\delta_{ij} \,.
\end{align}
The action in this gauge is more complicated and has been written in \cite{Handley:2019anl},
\begin{align}
S=\frac{1}{2} \int d^4 x \sqrt{\gamma} a^3\frac{\dot{\phi^2}}{H^2}\left[\left(\dot \zeta+ \frac{1}{a^2H}\frac{\zeta}{H}\right)^2-\frac{1}{a^2}\left(\dot \zeta+ \frac{1}{a^2H}\frac{\zeta}{H}\right)\left(\psi+\frac{\zeta}{H}\right)-\frac{1}{a^2}(\partial\zeta)^2+\frac{3}{a^2}\zeta^2
\right] \,.
\label{eq:actioncomovinggaguge}
\end{align}
Notice that $\dot\phi^2= \mpl^2(\dot H-a^{-2})$, hence in the action in comoving gauge reintroduces the dependence on $\mpl^2$.
\section{Solution of Lorentzian mode function}
\label{sec:solution_eq}
In this section we detail how to solve the equation for the massless wave equation in  in Lorentzian signature (as version of it in conformal time  is given in Eq.~\eqref{eq:Lorentizanequation}). We will write it as,
\begin{align}
\partial_t^2\varphi +3 \frac{\dot a}{a}\partial_t\varphi-\frac{1}{a^2}\nabla^2\varphi=0
\end{align}

To solve thi equation we will assume that the background is pure de Sitter, and we write it first in cosmic time where the scale factor is given  by  $a(t)=\frac{1}{H}\cosh (H t)$. The equation  becomes,
\begin{equation}
\partial^2_t\varphi+3H \tanh(H t)\partial_t\varphi-\frac{H^2}{\cosh^2(Ht)}\nabla^2\varphi=0 \,,
\end{equation}
 we then write the spatial part in terms of spherical harmonics,
\begin{equation}
\varphi(\xi,\Omega)=\sum_{p, l, m}\tilde\phi_{p, l, m} Y_{p, l, m}(\Omega) \,,
\end{equation}
where $\xi=H t$. The equation becomes,
\begin{equation}
\left[\partial_\xi^2+3\tanh\xi\partial_x+\frac{p(p+2)}{\cosh^2\xi}\right]\phi_p=0 \,.
\end{equation}
Using the substitution $z= \frac{1}{2}\left(1-i\sinh \xi\right)$ we have
\begin{equation}
\left[z(z-1)\partial^2_z-2(1-2z)\partial_z+\frac{p(p+2)}{4z(z-1)}\right]\phi_{p, l, m}=0 \,.
\end{equation}
Now writing $\phi=(-4z(1-z))^p v_{z,p}$ the equation becomes a Hypergeometric equation for $v_{z,p}$,
\begin{equation}
\left[z(1-z)\partial^2_z+(2+p-z(4+2p))\partial_z +p(3+p)\right]v_{z,p}=0
\end{equation}
The solutions are,
\begin{equation}
v_{z,p}=c_1 \,  _2F_1(p,3+p;2+p;z)+c_2(z)^{-1-2p} \, _2F_1(3,2;p;z)
\end{equation}
This solution is better  written in conformal time $a(\eta)=1/H \sec(\eta)$, where $\eta$ runs from $-\pi/2$ to $\pi/2$. We then have that,
\begin{equation}
\phi_{p}=c_1((1+p)\cos\ \eta+i\sin\ \eta )e^{-i(p+1)\eta}+c_2 ((1+p)\cos\ \eta-i\sin\ \eta )e^{+i(p+1)\eta}
\end{equation}
\section{Harmonics in the   flat space limit}
\label{sec:Harmonics}
Let us make the comparison to flat space more precise. First let us recall that the $S^3$ spherical harmonics can be written as\footnote{These formulae were derived by ~\cite{Lifshitz:1963ps,Harrison:1967zza} as quoted by \cite{Ratra:2017ezv}}
\begin{align}
Y_{p,l,m}=\sqrt{\frac{(l+1)\Gamma(p+l+2)}{\Gamma(p-l+1)}}\frac{1}{(\sin r)^{1/2}}P^{-l-1/2}_{p-1/2}(\cos r))Y_{lm}(\theta,\phi)
\end{align} where $Y_{mn}(\theta,\phi)$ are the usual $S^2$ spherical harmonics with $m\geq \vert n\vert$.
We would like to compute the asymptotic limit when $p$ is large. Expanding the Legendre polynomial for large $p$  and fixed $l$ we find that\footnote{Taken from https://dlmf.nist.gov/14.15},
\begin{align}
P^{-l-1/2}_{p-1/2}(\cos(r))=\frac{1}{l^{l+1/2}}\left(\frac{r}{\sin r}\right)^{1/2}(J_{l+1/2}(p r)+\mathcal{O}(1/p))
\end{align}
Now using the Stirling formula we have that for  large $l$  and fixed $m$,
\begin{align}
\sqrt{\frac{(p+1)\Gamma(p+l+2)}{\Gamma(p-l+1)}}=\sqrt{(p+1)l^{2l+1}(1+\mathcal{O}(1/p))}=p^{l+1}(1+\mathcal{O}(1/p))
\end{align}
Joining both pieces we find for large $p$ that ,
\begin{align}
Y_{p,l,m}=\frac{(pr)^{1/2}}{\sin r}J_{l+1/2}(l r)Y_{lm}(\theta,\phi)(1+\mathcal{O}(1/p)).
\end{align}
Notice that in the limit when $\sin r\approx r$ which corresponds to neglecting the curvature, the above formula simplifies to,
\begin{align}
Y_{p,l,m}\approx\sqrt{\frac{p}{r}}J_{l+1/2}(p r)Y_{lm}=\sqrt{\frac{2}{\pi}}p j_l(pr)Y_{lm}(\theta,\phi)
\end{align}
the last expression is the formula for the so called flat harmonics. These modes are orthogonal, indeed we have that,
\begin{align}
\int d^3 x \frac{2}{\pi} p p' j_l(p r)j_l' (p'r)Y_{lm}(\theta,\phi)Y^*_{l'm'}(\theta,\phi)&=\int r^2 dr d\Omega_2 \frac{2}{\pi} p p' j_l(p r)j_l' (p'r)Y_{lm}(\theta,\phi)Y^*_{l'm'}(\theta,\phi)\nonumber\\
&=\int dr r^2\frac{2}{\pi} pp'j_l(pr)j_l'(p'r)\delta_{l,l'}\delta_{m,m'}\nonumber\\
&=\delta(p-p')\delta_{l,l'}\delta_{m,m'}
\end{align} 
where in the second line we have used the orthogonality of the $S_2$ spherical harmonics and in the third the identity 
\begin{align}
\int_0^{\infty}x j_\alpha(ux)j_\alpha(vx)dx=\frac{\pi}{2u^2}\delta(u-v)
\end{align}
It will also be useful to derive the following relation,
\begin{align}
\int_0^{\infty}dl \frac{2l^2}{\pi} \sum_{lm}j_l(p r)j_{l}(pr')Y_{lm}(\Omega_2)Y^*_{lm}(\Omega_2')&=\int_0^{\infty}dp \frac{2p^2}{\pi} \sum_{l}\frac{2l+1}{4\pi}j_l(p r)j_l(p r')P_l\left(\frac{\vec{r}\cdot\vec{r}'}{rr'}\right)\nonumber\\
&=\int_0^{\infty}dp\frac{p^2}{2\pi^2}\frac{\sin (p\vert\vec{r}-\vec{r'}\vert)}{p\vert\vec{r}-\vec{r'}\vert}\nonumber\\
&=\delta(\vec{r}-\vec{r'}))
\end{align}
These are related to the usual Fourier modes trough the relation,
\begin{align}
e^{i\vec{p}\cdot\vec{x}}=\sum_{lm}i^l j_l(p r)Y^*_{lm}(\theta,\phi)Y_{lm}(\theta,\phi)
\label{eq:SphericalHarmonics_rel1}
\end{align}
wher we have written $p$ as a vector. 
Using this expansion we can obtain an expression for $\phi_{p,l,m}$ in term of the Fourier coefficients. To do so let us write,
\begin{align}
\phi(\vec{x})&=\int\frac{ d^3\vec{p}}{(2\pi)^3}e^{i\vec{p}\vec{x}}\phi_p\nonumber\\
&=\int _0^{\infty}dp \sqrt{\frac{2}{\pi}}p\sum_{lm}j_l(pr)Y_{lm}(\theta,\phi)\phi_{plm}
\end{align}
Using Eq.~\eqref{eq:SphericalHarmonics_rel1} we can write
\begin{align}
\phi^*_{plm}=\frac{i^p}{(2\pi)^{3/2}p}\int d\Omega Y_{lm}(\Omega)\phi_p
\end{align}
and
\begin{align}
\phi_p=i^{-p} Y_{lm}(\theta,\phi)\phi^*_{plm}
\end{align}
It will also be useful to write, 
\begin{align}
i^p\frac{p}{(2\pi)^{3/2}}\int d\Omega Y_{lm}(\Omega)e^{-i\vec{p}\cdot\vec{x}}=\sqrt{\frac{2}{\pi}}j_p(pr)Y_{lm}(\Omega)
\end{align}
Now the two point correlation function is given by,
\begin{align}
\langle\phi_{p,l,m}\phi^*_{p',m',l'}\rangle&=\left(i^l\frac{p}{(2\pi)^{3/2}}\int d\Omega Y_{lm}(\Omega)\right)\left(-i)^{p'}\frac{p'}{(2\pi)^{3/2}}\int d\Omega' Y_{l'm'}(\Omega')\right)(2\pi)^3\delta(\vec{p}-\vec{p}')P(p)\rangle\nonumber\\
&=\delta(p-p')\delta_{l,l'}\delta_{m,m'}P(k)
\end{align}
where $\langle\phi_{\vec{p}}\phi_{\vec{p}'}\rangle=(2\pi)^3\delta(\vec{p}-\vec{p'})P(p)$
In the same way it is possible to obtain an analogous expression for the three point function
\begin{align}
\langle\phi_{p,l,m}\phi_{p',l',m'}\phi_{p'',l'',m''}\rangle&=G^{p,p',p''}_{l,l',l'';m,m',m''}B(p,p',p'')
\end{align}
where $B(p,p',p'')$ is the bispectrum in momentum space defined through,
\begin{align}
\langle\phi_{\vec{p}}\phi_{\vec{p}'}\phi_{\vec{p}''}\rangle=(2\pi)^3\delta(\vec{p}+\vec{p}'+\vec{l}'')B(p,p',p'')
\label{eq:relation_delta_G}
\end{align}

\bibliographystyle{JHEP}
\bibliography{closed_universe_NG}

\end{document}